\documentclass[a4paper,12pt]{article}
\usepackage{amsmath}
\usepackage{amsfonts}
\usepackage{amssymb}
\usepackage{epsfig}
\usepackage{bbm}
\usepackage{latexsym}
\usepackage{graphicx}
\usepackage{fancybox}
\usepackage{mathrsfs}

\newcommand{\p} \prime
\newcommand{\e} \epsilon
\newcommand{\la} \lambda 

\newcommand{\LV}{\Lambda_{LV}}
\newcommand{\sign} {\rm sign}
\newcommand{\llv} {\Lambda_{LV}}
\newcommand{\om} \omega   \newcommand{\Om} \Omega

\newcommand{\al} \alpha
\newcommand{\bt} \beta
\newcommand{\be} {\begin{equation}}
\newcommand{\ee} {\end{equation}}
\newcommand{\ba} {\begin{eqnarray}}
\newcommand{\ea} {\end{eqnarray}}
\def\lrD{\mathrel{{\cal D}\kern-1.em\raise1.75ex\hbox{$\leftrightarrow$}}}
\def\lr #1{\mathrel{#1\kern-1.25em\raise1.75ex\hbox{$\leftrightarrow$}}}

\begin{document}

\vskip 1cm
\begin{center}
   {\large  \sc 
   Confronting the trans-Planckian question\\of inflationary 
   cosmology\\
   
   with dissipative effects
   }
\end{center}

\vskip 1cm

\begin{center}
       Renaud Parentani 
\end{center}

\vskip 1cm

\begin{center}
{\it Laboratoire de Physique Th\'eorique, UMR 8627},
\\ B\^at. 210, Universit\'e Paris-Sud 11,
 \\ 91405 Orsay Cedex, France
\end{center}

\vskip 1. truecm

\vskip .3 truecm
%

\begin{abstract}
We provide a class of QFTs which exhibit dissipation 
above a threshold energy, thereby breaking Lorentz invariance. 
Unitarity is preserved by coupling the 
fields to 
additional degrees of freedom (heavy fields) which introduce the rest frame. 
Using the Equivalence Principle, we define these theories in arbitrary curved spacetime.
We then confront the trans-Planckian question of inflationary cosmology.
When 
dissipation increases with the energy,
the quantum field describing adiabatic perturbations is completely damped
at the onset of inflation. 
However it still exists as 
a composite operator made with the 
additional fields. 
And 
when these are in their ground state, the standard power spectrum obtains
 if the threshold energy is much larger that the Hubble parameter.
In fact, as the energy redshifts below the threshold,
 the composite operator behaves as if it were a 
free field 
endowed with standard vacuum fluctuations. 
 The relationship between our models and the Brane World scenarios
 studied by Libanov and Rubakov displaying similar effects is discussed. 
 The signatures of dissipation will be studied in a forthcoming paper.

%
\end{abstract}






\newpage

\section{Introduction and presentation of the settings}

 Even though relativistic QFT provides an excellent description of
 particle physics, being non-compact, Lorentz symmetry 
can only be tested up to a certain energy scale. 
Thus one cannot exclude that some unknown high energy processes 
break the invariance under boosts,
thereby introducing a threshold energy $\LV$,
and a preferred
frame.
It is therefore of interest to determine 
what would be the 
signatures when 
this possibility
is realized~\cite{Jacobson:2005bg}.

This question is particularly relevant for inflationary cosmology
because primordial density fluctuations arise from 
vacuum 
fluctuations which had very short wave lengths
(very large proper frequencies) at the onset of inflation\cite{Mukhanov:1990me}. 
Indeed, 
the initial
frequencies of the modes we observe today in the CMB anisotropies
were all larger than 
\be
\Omega_{in} = H \, e^{N_{extra}}\, ,
\label{Nex}
\ee
where $H$ is the value of the Hubble parameter during inflation, 
and where $N_{extra}$ is the number of e-folds from the onset 
of inflation till $t_0$, the moment when the (comoving) scale of our visible
universe exited the Hubble radius. (The total number of 
e-folds is thus $N_0 + N_{extra}$ where $N_0 \sim 60$ is the 
number of e-folds 
from $t_0$ 
till the end of inflation.) 
Irrespectively of 
the inflationary scenario, 
$\Omega_{in}$ is thus larger than the Planck mass $M_{Pl}$ 
when $N_{extra}> 
\ln M_{Pl}/H$. 
Since $H$ should be 
of the order of $10^{-5} M_{Pl}$, 
the initial frequencies of observable modes were all trans-Planckian 
when $N_{extra} > 
5 \ln 10\sim 12$.
Moreover, since
 in most scenarios 
$N_{extra} \gg 12$, 
$\Omega_{in}$
was generically much larger than the Planck mass.

In the absence of  Quantum Gravity, 
there is no understanding of the nature of the degrees of freedom
at these scales:
we neither know what is their Hilbert space,
nor how they propagate, and this, because 
the notions of differential geometry
used when describing (quantum) fields 
in expanding universes 
might not apply at very short distances.
In particular 
eq. (\ref{Nex}) which is based
on a relativistic 
mode equation in an accelerating universe
might loose its validity above a certain Ultra-Violet 
scale $\Lambda_{LV}$.  
Since this 
scale might significantly differ
from the Planck mass, we shall keep it 
distinct from the Planck mass.
We shall also use 
the abbreviations $LI$
and $LV$ for Lorentz Invariance and Lorentz Violation respectively. 

In this context, it is instructive
 to phenomenologically parameterize the 
 deviations from 
$LI$ 
and determine their signatures 
on 
inflationary spectra. 
This line of thought has been 
proposed
in the 
context of Hawking radiation from black holes, wherein 
the asymptotic quanta also arise from vacuum
fluctuations with exponentially growing frequencies~\cite{Unruh:1994je,Jacobson:1993hn}.
 The deviations from $LI$ 
 have been characterized by  non-linear dispersion relations
  imposed in a particular rest frame,
 \be
 \Omega^2 = F_n(p^2) = p^2 \, ( 1 \pm p^{2n} /\Lambda_{LV}^{2n}
  + O(p^{2n+2}))   ,
 \label{disprel}\ee
 where $\Omega$ and $p$ 
 are respectively the frequency and the norm of the spatial momentum
 measured in that frame. 
 The first deviation
 is characterized by a power of $p/\llv$ and the sign defining 
  super-luminous $(+)$ and sub-luminous $(-)$ cases.  
 When 
  $\Lambda_{LV}$ is 
 much larger than Hawking temperature, and when the frame
 is freely falling, it was shown that the asymptotic 
 properties of Hawking radiation are unmodified, 
 even though the near horizon propagation is radically modified 
 when $\Omega$ is larger than $\llv$. 
 The 
 robustness of asymptotic properties  relies on the 
 adiabatic character of the evolution of the vacuum 
 (ground) state~\cite{Brout:1995wp}.
The same program was then applied to the inflationary 
spectra~\cite{Martin:2000xs,Niemeyer:2000eh},
and  similar results were obtained because, 
in this case as well, the ground state adiabatically evolves 
when $\sigma= H/\Lambda_{LV} \ll 1$~\cite{Niemeyer:2001qe}. 
However, so far, 
 only dispersive models have been systematically studied.

The aim 
of the present work is to extend this analysis to
dissipative models.
To this end, we shall first construct QFT displaying dissipative effects
in the UV. 
Indeed, 
unlike dispersion, dissipation requires 
 enlarged dynamical settings 
because
 if one tries to introduce dissipation
from the outset in eq. (\ref{disprel}), 
one looses both 
unitarity and
predictability.
To preserve them, 
we shall work with Hamiltonian theories in which dissipative effects
are caused by interactions with additional fields. 
Doing so, we shall establish that dissipative effects are {\it generic}. 
That is, when starting with 
a Lagrangian in which $LI$ is broken 
in the UV, the effective theory 
unavoidably
develops dissipation above a certain energy scale, simply because nothing
can prevent this. (With relativistic QFT instead, $LI$ did prevent it).
 
Since we want to construct generalized QFT, we should decide
 what to keep.
First, we want to 
preserve the unitarity of the description because
the calculation of the power spectrum 
of (adiabatic) 
density fluctuations \cite{Mukhanov:1990me} rests on the identification
of a scalar field, 
hereafter called $\phi$, 
which obeys Equal Time Commutators (ETC). 
This identification is necessary 
to fix 
the rms 
amplitude of density fluctuations in the vacuum. 
Explicitely 
  the power spectrum 
  is given \ba
 {P}_p(t) &\equiv & 4 \pi p^3 
 \int \Big({dx \over 2 \pi}\Big)^3 \, e^{i {\bf p x}} \,  G_{a}(t,{\bf x}; t, {\bf 0}),
 \label{PandS}\ea
 where $G_a$ is the anti-commutator of 
 $\phi$ evaluated in the 
 asymptotic (Bunch-Davies) vacuum 
 much 
 after horizon exit. 
In the absence of dissipation,
$P_p$ can be deduced from the norm
of $\phi^{in}_p(t)$, 
the Fourier mode of $\phi$ with asymptotic positive frequency. 
However
in the presence of dissipation, 
this notion of (free) mode disappears. 
Hence the knowledge of $G_a$ 
 becomes {necessary} since eq. (\ref{PandS}) gives
the only way to obtain the power spectrum. 
 
To get dissipation
we shall 
thus introduce 
additional 
degrees of freedom, hereafter called $\Psi$.
Then 
the whole system 
($\phi$ + 
$\Psi$) will evolve unitarily, 
by construction. 
This guarantees that the ETC of 
$\phi$
is exactly 
preserved (in a non trivial way since $\phi$ undergoes dissipation).
Moreover, 
the (dressed) 2pt 
function of $\phi$ will
be given by the usual QM trace
\be
G_W(x,y)= {\rm AR}
\, \Big[  \hat \rho_T \, \hat \phi(x) \, \hat \phi(y)  \big],
\label{Gw}\ee
where $\hat \rho_T$ is the initial 
matrix density of the total 
system, 
where 
$\hat \phi(x)$ is the Heisenberg field operator evolved with the 
time ordered exponential of 
the total Hamiltonian, 
and where the trace is taken over both $\Psi$ and $\phi$.
The anti-commutator $G_a$ determining the power spectrum in eq. (\ref{PandS})
is simply the symmetric part of $G_W$. 


Our second requirement concerns the properties of dissipative effects.
When considering the theory in vacuum and in Minkowski space-time, 
we impose that these 
effects 
preserve  the stationary, the homogeneity, and
isotropy of flat space-time. These requirements
define a preferred frame 
which is inertial and globally defined.
Then irrespectively of the choice of 
$\Psi$
and $\Psi$-$\phi$ interactions, 
the Fourier transform of the retarded 
Green function, $G_r(x,y)= \theta(t_x - t_y)\, 2$Im$G_W(x,y)$, 
is of the form
\be
G_r(\om,\vec p) = 
{ 
1\over \Big(-\om^2 + p^2 + \Sigma_r(\om,p)  \Big)} \, .
\label{GfF}
\ee
In 
the true vacuum, at the level of the 2 pt functions,
the dissipative (dispersive) effects 
 are thus completely characterized by the imaginary and odd part in $\om$ 
 (real and even part) of 
 the self energy $\Sigma_r(\om,p)$. 
 
 In these expressions, the energy $\om$ and the momentum
 square $p^2$ are defined in the preferred frame. 
%
  The novelty 
 is that $\Sigma_r$ is  a function of $\om$ and $p$ separately,
 and not only a function of the relativistic invariant $\om^2 - p^2$
 as it is the 
  case in $LI$ QFT. 
 Therefore dissipation can become significant above a critical
 energy {\it on the mass shell}, i.e. along the 
 minima of the denominator of eq. (\ref{GfF}); 
 a possibility forbidden in $LI$ theories.
 For instance, one verifies that 
  the following self-energies induce
 significant dissipation only above $\LV$
\be
{\rm Im}\Sigma^{(n)}_r(\om, p) 
    = - {\om \over \llv}
    \, p^2 
   \left({p^2 \over \llv^2}\right)^{n} .
\label{nS}\ee
With this equation we have 
identified the relevant (lowest order) 
quantity governing dissipation, the equivalent of the 
first order deviation in eq. (\ref{disprel}). 
In the body of the paper, we shall provide 
Lagrangians of $\Psi$ and $\phi$ 
which give 
this class of self-energies labeled by $n$.
Rather than focusing on a 
particular case, we shall 
describe the whole class 
of 
 dissipative behaviors 
(at the level of 2-point functions).
%
%
We shall thus follow a phenomenological approach based on 
{\it minimal}
assumptions (i.e. unitarity), 
rather than 
some
"inspired" approach
(e.g. by string theory, or branes scenarios~\cite{Libanov:2005yf,Libanov:2005nv}) 
 which would 
 lead 
to a particular sub-class of models.

Our third requirement concerns the extension of our models
from Minkowski space to curved space-times.
To define our QFT in curved space, we simply implement
the Equivalence Principle (EP). It 
 fixes the action density  our models to be the
%
%
covariantized version of that we had in 
Minkowski space (up to some non-minimal 
 coupling).
To perform the covariantization,
 it is useful to    characterize the preferred frame 
 in a coordinate invariant way by $l^\mu$, a unit time-like 
 vector field~\cite{Jacobson:1996zs}. 
  In terms of this vector, 
  $\om$ and $p^2$  are given by 
  \be
  \om \equiv   l^\mu \, p_\mu \, , 
  \quad  p^2 \equiv  \perp^{\mu\nu} \, p_\mu p_\nu \ ,
  \label{covlmu}\ee
  where $\perp^{\mu\nu} \equiv g^{\mu\nu} + l^\mu l^\nu$ is the 
  (positive definite) metric
  in the spatial sections orthogonal to $l^\mu$.   
The covariant action
will be a sum of scalar functions
of the four local fields $\phi$, $\Psi$, $g_{\mu \nu}$ and $l^\nu$
which reduces to the Minkowski model in the zero curvature limit.
Very 
importantly for us, we shall see that the EP
guarantees the adiabaticity of the evolution of the 
(interacting) ground state as long the gradients of the metric are much smaller
than the UV scale $\LV$. 
 
\vskip .2 truecm


For the interested reader, we mention 
that 
additional comments which place the
present work in broader contexts can be found 
in the introduction of~\cite{Parentani:2007uq}. These comments concern Quantum Gravity 
and the description of ''mode creation'' in expanding universes.

\section{Dissipation in Minkowski space from $LV$ effects}

In this Section, we provide 
a class of models defined in Minkowski
spacetime which exhibit dissipative effects above
a certain energy scale $\llv$. Stationarity, homogeneity
and isotropy will be exactly preserved. Therefore, 
the only invariance
of relativist QFT which is 
broken is that under boosts.
These theories define a preferred rest frame which is globally defined, 
as it is the case 
in FLRW space-times. 
For the simplicity of the presentation, we shall first 
work in the preferred frame. At the end of this Section we shall
covariantize our action and generalize it to curved spacetime.

\subsection{Free field settings}

We start with a brief presentation of the free field quantization 
to introduce notations and to point out what are the 
properties which will be lost in the presence of dissipation.

The action of our free massless field $\phi$ is the usual one:
\be
S_\phi = {1 \over 2} \int dt d^3{x} \, (\partial_t\phi^2 -
 \partial_{\bf x}\phi \cdot  \partial_{\bf x}\phi  ) \, ,
 \label{freeS}
\ee 
where $t$ and ${\bf x}$ are 
Cartesian coordinates.
Due to the homogeneity of space,
the equation of motion can be analyzed mode by mode:
\be
\phi(t,{\bf x}) = \int\! {d^3{p}\over (2 \pi)^{3/2}}\,  e^{i\bf p \cdot x} \phi_{\bf p}(t) 
\, .
\ee
The Fourier mode 
$\phi_{\bf p}(t)$ obeys
\be
(\partial^2_t + \om_p^2 ) \phi_{\bf p} =0\, ,
\label{eqom}
\ee
where $\om_p^2= p^2 = \bf p \cdot p$ is the standard relativistic dispersion relation.
Notice that 
eq. (\ref{eqom}) is second order, homogeneous (no source term),
and time reversible (no odd power of $\partial_t$), three
properties we shall loose when introducing
interactions breaking $LI$.

In homogeneous space-times, 
the Equal Time Commutator between the Heisenberg 
field operator and its momentum
implies that 
the mode 
operator (q-number) $\phi_{\bf p}(t)$ 
obeys 
\be
[\phi_{\bf p}(t), \partial_t \phi^\dagger_{\bf p'}(t)]
 = i \, \delta^3({\bf p}-{\bf p'}) \, .
\label{etc}
\ee
When decomposing this operator as
\be
\phi_{\bf p}(t) = a_{\bf p} \, \phi_p(t) + a^\dagger_{-\bf p} \, \phi_p^*(t)\, ,
\ee
 where the destruction and creation
 operators satisfy the usual commutators
 \be
 [a_{\bf p}, a^\dagger_{\bf p'}]= \delta^3({\bf p}-{\bf p'})  
 \, , \quad [a_{\bf p}, a_{\bf p'}]=0 \, ,
 \label{uscom}
 \ee
 eq. (\ref{etc}) 
 is verified 
because the Wronskian of the positive frequency
(c-number)
 mode \be
\phi_p(t) = e^{- i \om_p t}
/(2 \om_p)^{1/2} \, ,
\label{fm}
\ee
is constant (and conventionally taken to be unity). 

Had an odd term like $\gamma \partial_t$ be present in eq. (\ref{eqom})
the constancy of the Wronskian would have been lost.
Hence
the possibility of realizing the ETC (\ref{etc})
with the help of eq. (\ref{uscom}) would have been lost as well. 
This already indicates
that, unlike dispersive (real) effects,
dissipative effects
require more general settings than the above. 

\subsection{Interacting models breaking $LI$, general properties}

We now introduce additional 
degrees of freedom, 
here after collectively named ${\bf \Psi}$,
which 
induce dissipation above the energy $\llv$. 
We shall work with a particular class of 
models 
in order to get an 
exact (non-perturbative) 
expression 
for the two-point function of
eq. (\ref{Gw}). 
Before introducing these 
models,
we derive 
general results 
valid for all  unitary QFT's 
possessing
dissipative effects above $\llv$ in the ground state (the interacting vacuum). 

We assume that the total action decomposes as
\be
S_T = S_\phi +  S_{{\bf \Psi}} +  S_{\phi, {\bf \Psi}} \, ,
\ee
where the first action is that of eq. (\ref{freeS}), the second one
governs the evolution of the ${\bf \Psi}$ fields, and the last one
the coupling between $\phi$ and these new fields. 
We also impose that the last two actions preserve the homogeneity
and isotropy of Minkowski space but break the 
invariance under boosts. %
%
%
From now on, the 
coordinates $t, {\bf x}$ 
are at rest with respect to the preferred frame defined by 
$S_{{\bf \Psi}} + S_{\phi, {\bf \Psi}}$,
i.e.,  $\partial_t \equiv l^\mu \partial_\mu$
in the covariant notation of eq. (\ref{covlmu}). 

When the state of such system is
homogeneous, 
the Fourier transform of the 2pt 
function of eq. (\ref{Gw})
is of the form
\ba
G_{\bf p, \bf p'}(t,t') 
 &=& {\rm Tr}\, [ \hat \rho_T \, \hat \phi_{\bf p}(t) \, \hat 
 \phi^\dagger_{\bf p'}(t')] \, ,
 \nonumber \\
&=& G_W(t,t'; p) \, \delta^3({\bf p}-{\bf p'}) \, .
\label{Gf}
\ea
At this point, an 
important remark should be made. 
In the presence of interactions, 
 the notion (and the usefulness) of the time-dependent 
 modes of eq. (\ref{fm})
 has disappeared
 whereas the  
 function $G_W(t,t'; p)$ of eq. (\ref{Gf}) is always 
 well-defined, 
  for all choices of ${\bf \Psi}$ 
  and $S_{{\bf \Psi}} +  S_{\phi, {\bf \Psi}}$.  

When the situation
is stationary,
$ G_W(p; t,t')$ further simplifies in the frequency
representation:
\be
G_W(t,t'; p) = \int {d\om \over 2\pi} e^{-i\om(t-t')} G_W(\om, p) \, .
\label{FTdef}\ee
%
When working in the ground state, one finds 
the simplest case,
because whatever the ${\bf \Psi}$
fields may be, the Fourier transform of the
time ordered (Feynman) propagator, 
\be
2 G_F = {\rm Re}G_W + i\,  {\rm Im} G_{W} \, {\sign}(t-t'), 
\ee 
is always of the same form as $G_r$ 
in eq. (\ref{GfF}),
and therefore characterized by a {single} (complex)
function $\Sigma_F(\om,p)$. 
When restricting attention to Gaussian models,
$\Sigma_F(\om,p)$ is given by a 1-loop 
calculation, whereas in general, it contains a series of 1PI graphs. 

In non-vacuum states and in non-stationary situations, 
the 2pt functions and self-energies have 
a more complicated structure~\cite{Arteaga:2007us}.
To 
understand 
this structure it is useful to study separately  
the commutator 
 $G_c = i {\rm Im} G_W$
 and the anti-commutator $G_a = {\rm Re} G_W$.
%

Let us conclude with two remarks.
First, the effective dispersion relation of $\phi$ 
is {\it a posteriori} defined by the poles of eq. (\ref{GfF})
which are governed by $\Sigma(\om,p)$~\cite{Arteaga:2003we,Parentani:2007mb}.
In this way, non-trivial dispersion relations arise from
dynamical processes rather than from being introduced from
the outset. The present work therefore provides physical 
foundations (and restrictions, as later discussed) 
to the kinematical approach which is usually adopted~\cite{Jacobson:2005bg}. 
Second, from analyzing dynamical models, we shall see
that, even in the vacuum, 
{\it on-shell} dissipative effects (i.e. dissipation arising along the
minima of the denominator of eq. (\ref{GfF}))
are 
unavoidable
when $LI$ is broken in the UV by the action $S_{\bf \Psi}+S_{\phi, \bf \Psi}$, 
in complete opposition 
with the fact that 
on-shell dissipation is forbidden when 
working (in the vacuum) with  
$LI$ actions. 

\subsection{Gaussian models}

To simplify the calculation of $\Sigma(\om,p)$
and to get non perturbative expressions, 
we assume that the action $S_T$ 
is quadratic in all field variables.
At first sight, this could be considered as an 
artificial hypothesis. However,
it should be recalled that we are {not} 
after computing $\Sigma$ from first principles. Rather we aim to 
compute the signatures of
the power spectrum (\ref{PandS}) 
{\it given} the properties of $\Sigma$, 
following the 
approach adopted in 
\cite{Unruh:1994je,Brout:1995wp,Martin:2000xs,Niemeyer:2000eh}.

%
%
Given that we are preserving the homogeneity of Minkowski space, 
the Gaussian assumption 
implies that 
the total action splits as 
\be
S_T = \int \! d^3p \, S_T({\bf p}) \, , 
\ee
where each action $ S_T({\bf p})$
depends only on $\phi_{\bf p}$ and the ${\bf p}$-th
Fourier component of ${\bf \Psi}$.
The structure of these actions is 
\ba
 S_T({\bf p}) &=& {1 \over 2 } \int dt\ \phi_{\bf p}^* 
 (- \partial_t^2 -  \om_p^2 ) \phi_{\bf p} 
+ {1 \over 2 } \Sigma_i \int dt\ \Psi^*_i({\bf p})
 ( - \partial_t^2 - \Omega^2_i(p)) \Psi_i({\bf p})
\nonumber\\
&& + \Sigma_i  \int dt\ g_i(p) \,  \phi_{\bf p} \, 
\Psi^*_i({\bf p}) \, ,
\label{Sd}
 \ea
 where $i$ is a discrete (or continuous) index, 
 where $\Omega_i(p)$ is the energy of the quanta of the
 oscillators $\Psi_i({\bf p})$, 
 and where $g_i(p)$ is the coupling constant at fixed $p, i$. Since
$\phi_{\bf p}^\dagger = \phi_{\bf - p}$, $\Psi_i^\dagger({\bf p})
= \Psi_i({\bf -p})$,
 $S_T({\bf p}) + S_T({\bf -p})$
is real. Instead of choosing a priori one model, 
it is more instructive to solve the equations of motion
 without specifying 
the set of $\Psi_i({\bf p})$,
their energy $\Omega_i(p)$ and their coupling $g_i(p)$. 
We shall choose them in due course.
To preserve stationarity in Minkowski space, the $\Omega$'s  and  the $g$'s
must be time independent. 
 When 
 $\Omega_i^2(p) \neq M_i^2 + p^2$, the kinetic action of $\Psi_i({\bf p})$
 breaks 
 $LI$ and defines 
 the preferred frame. On the contrary when  $\Omega_i^2(p) = M_i^2 + p^2$
 the preferred frame is only defined by $S_{\phi \Psi}$
  through the $p$-dependence of the coupling functions $g_i(p)$. 

Models of this type have been used 
for different purposes. They have been introduced 
(in their continuous version) 
to study non-pertubatively atomic transitions
 see \cite{1976AnPhy..98..264A}
 and refs. therein, 
 see also \cite{Raine:1991kc,Massar:1900vg} for an application to
 the Unruh effect. They have been used in Quantum Optics~\cite{QNoise},
  to model quantum Brownian motion,
%
 and to study decoherence 
 effects~\cite{Unruh:1989dd}.
 Depending on the aims, 
 they can be
 solved and analyzed by means of different methods.
 In what follows we shall 
 use the simplest approach based on Heisenberg picture.\footnote{
 Even though legitimate, we shall not use the general methods (Influence Functional, Master Equation, Closed Time Path Integral)~\cite{QNoise} 
 which have been developed to study 
"open quantum systems" 
because they  somehow 
 hide the simplicity of the present models. 
  Moreover 
 we are planning (in a subsequent work)
 to 
 study the correlations between $\phi$ and $\Psi_i$. Therefore we shall
 treat $\phi$ and $\Psi_i$ on equal footing 
 as in \cite{1976AnPhy..98..264A,Raine:1991kc,Massar:1900vg}.}
Since we shall work at zero temperature, 
 this approach offers a simple characterization of the state of the system 
in terms of the ground states of the free modes before $\phi-\Psi$ interactions are turned on.

 
 
To prepare the application to inflation, 
  we shall
  treat $\omega_p^2$, $\Omega_i^2$ and $g_i$ as arbitrary functions
  of time (in cosmology these quantities  become time
  dependent through their dependence in the scale factor $a(t)$). 
 The equations of motion are
 \ba
 &&( \partial_t^2 +  \omega_p^2 ) \, \phi_{\bf p} = \Sigma_i \,  g_i(p) \, 
 \Psi_i({\bf p}) \, ,
 \label{eqphi}
 \\
 && (  \partial_t^2 +  \Omega^2_i) \Psi_i({\bf p}) = 
 g_i(p)  \phi_{\bf p} \, .
 \label{eqPsi}
 \ea
 The general solution of the second equation reads
 \be
  \Psi_i({\bf p}, t)=\Psi_i^o({\bf p}, t) + \int dt' R_i^o(t,t'; p) \,
  g_i(t'; p) \, \phi_{\bf p}(t')\, ,
  \label{solPsi}
  \ee
  where $\Psi_i^o({\bf p}, t)$ is a free solution which depends
  on initial conditions imposed on $\Psi_i({\bf p})$. 
  The second term contains
  $R_i^o(t,t';p) $, the (free) retarded Green function 
  of $\Psi_i({\bf p})$. It obeys
  \be
  (  \partial_t^2 + \Omega^2_i(p)) \,  R_i^o(t,t'; p) = \delta(t-t')\, ,
  \ee
  and vanishes for $t < t'$.
  Injecting eq. (\ref{solPsi}) in eq.  (\ref{eqphi}) one gets
  \be
  ( \partial_t^2 + \om_p^2 ) \phi_{\bf p} = 
  \Sigma_i \, g_i(t; p) 
  \Psi_i^o({\bf p}, t)
  + \Sigma_i   \, g_i(t; p) \int dt' 
  R_i^o(t,t'; p) 
   g_i(t'; p)  \phi_{\bf p}(t') \, .
  \label{eqtosimpl}\ee
  The 
  solution of this equation has always the following structure 
  \be
  \phi_{\bf p}(t) = \phi_{\bf p}^d(t) + \int dt' G_r(t,t'; p) 
  [\Sigma_i \,  g_i(t'; p)
  \Psi^o_i({\bf p}, t')] \, .
  \label{trues}
  \ee
  The first term is the ''decaying'' solution. It contains
  all the information about the initial condition of $\phi_{\bf p}$.
  The second term is the ''driven'' solution.
  It is governed by the initial conditions of $\Psi_i({\bf p})$
  and by the (dressed) retarded Green function, the solution of
  \be
  \int\!dt' [ \, \delta(t-t')(\partial_{t'}^2 + \om_p^2) 
  - \Sigma_i  \, g_i(t; p) \,
  R_i^o(t,t'; p) \,
  g_i(t'; p) ] \, G_r(t',t_1; p)  = \delta(t-t_1)\, . 
  \label{Gr}
  \ee
  Notice that $\phi_{\bf p}^d(t')$ is an 
   homogeneous solution of this equation.
  Therefore the evolution of both $\phi^d$ and $G_r$
  fully takes into account, through the non-local term in the 
  above bracket, the back-reaction due to the coupling
  to the $\Psi_i$. 
  In Gaussian models,
  it is quadratic in $g_i$. Hence 
   $\phi^d$ and $G_r$ 
  are series containing all powers of $g_i$. 
   Moreover 
  since $g_i(t;p)$
  are arbitrary functions of $p$ and $t$,  at this point, 
  there is no reason to consider non-Gaussian models.
 
  To conclude this subsection, 
we notice that
  eq. (\ref{trues}) {also} furnishes the exact solution
  for the (Heisenberg) mode operator $\phi_{\bf p}(t)$
  because the equations we solved were all linear.
  Since the power spectrum in inflation is obtained from 
  vacuum fluctuations, 
  instead of further analyzing the time dependence 
  of the mode 
  (as one would do in classical terms), 
  it is more 
  relevant to study 
  the correlation functions of $\phi_{\bf p}(t)$.
  
 \subsection{Structure of two-point correlation functions}
 
 This sub-section mainly contains well-known results
 which follow from the linearity of eq. (\ref{trues}).
 The key result we shall later use 
is given in eq. (\ref{i1}).
 
 Since our models are Gaussian, 
  {the} 2pt function of eq. (\ref{Gf}) 
  governs {\it all} observables 
  built with the Heisenberg field operator $\phi$.
  To analyse it, 
  as already mentioned, 
  it is appropriate to study separately the
  commutator and the anti-commutator. 
  
  We start with the simple part, the commutator
  \be
  G_c(t,t';p)\, \delta^3({\bf p}-{\bf p}') 
  \equiv {\rm Tr}[ \rho_T \, [\phi_{\bf p}(t) , \phi^\dagger_{\bf p'}(t')]_-\, ]\, .
  \label{Gct}
\ee
From eq. (\ref{trues}) one sees that 
it decomposes into two terms, one due to the 
non-commuting character of $\phi^d$, the other 
due to that of $\Psi_i^0$. In addition, 
since both 
commutators are c-numbers, it is independent
of $\rho_T$, the state of the system. Hence, for all Gaussian models, one has
\be
G_c(t,t';p) =  [\phi^d(t) , \phi^d(t')]_-\, 
+ \int\!\!\!\!\int\!dt_1 dt_2 \,  G_r(t, t_1) G_r(t',t_2) D(t_1,t_2)\, ,
\label{Gcgs}
\ee
where the ''dissipative'' kernel $D(t_1,t_2)$ is given by
\be
D(t_1,t_2)=
\Sigma_i \Sigma_j \, g_i(t_1) \,  g_j(t_2) \,  [\Psi_i^o(t_1) ,\Psi_j^o(t_2)]_- 
= \Sigma_i \, g_i(t_1) \, G_{c, i}^o(t_1, t_2)\,  g_i(t_2) \, .
\label{d}
\ee
Notice how this kernel combines the 
coupling $g_i$ 
and the non-commuting properties of $\Psi_i$. 

The next 
property of $G_c$ is more relevant. 
To all orders in $g_i$ and
for all 
sets of $g_i, \Omega_i$ (even with arbitrary
time dependence), one obtains
\be
i \partial_t G_c(t,t';p)\vert_{t=t'} = 1 \, .
\label{i1}
\ee
This identity 
corresponds to the ETC of eq. (\ref{etc}). 
The $1$ on the $rhs$ is guaranteed by the
Hamiltonian character of the evolution of the entire system $\phi + \Psi$.
It is therefore this equation which replaces the constancy of 
the Wronskian 
that was relevant in the case of free evolution. 

Eq. (\ref{i1}) is crucial for us for two reasons.
First, since 
the operator $\phi^d(t)$ in eq. (\ref{trues})
decays in $t-t_{in}$ where
$t_{in}$ is the moment when the interactions are turn on
--because it is an homogeneous solution of eq. (\ref{Gr})--
 the first term in eq. (\ref{Gcgs}) decays
as $\exp-\gamma(t+t'-2t_{in})$. Therefore 
 when $\gamma(t+t'-2t_{in}) \gg 1$,
 the non-commuting properties
 $ \phi_{\bf p}$ are {\it entirely
due} to those of the environment degrees of freedom, $ \Psi^o_i$. 
Secondly, these 
sum up exactly to $1$, as if the driven term of eq. (\ref{trues}) were a canonical degree of freedom.

We now analyze the anti-commutator, 
\be
G_a(t,t';p)\, \delta^3({\bf p}-{\bf p}') 
  \equiv {\rm Tr}[ \rho_T \, \{\phi_{\bf p}(t) , 
  \phi^\dagger_{\bf p'}(t')\}_+\, ]\, .
\label{Gags0}
\ee
When 
the (initial)
density matrix factorizes, $\rho_T = \rho_\phi \,  \rho_\Psi$,
as it is the case in the "free" vacuum before the interactions are turned on, 
$G_a$ also splits into two terms,
\be
G_a(t,t';p) =  {\rm Tr}[ \rho_\phi \{\phi^d(t) \,, \phi^d(t')\}_+\,] 
+ \int \!\!\!\!\int\! dt_1 dt_2 \, G_r(t, t_1) G_r(t',t_2) N(t_1,t_2)\, .
\label{Gags}
\ee
The first term only depends on the initial state of $\phi$. 
Similarly, 
the driven term only depends on the state of the environment
through the ''noise'' kernel
\be
N(t_1,t_2)=
\Sigma_i \Sigma_j  {\rm Tr}[ \rho_\Psi \,
 \{ g_i(t_1)\Psi_i^o(t_1) \,, g_j(t_2)\Psi_j^o(t_2)\}_+\,]  \, .
 \label{n}
\ee
As for the commutator, in the presence of dissipation, 
the first term exponentially decays, expressing the
progressive erasing on the information contained in the initial state
of $\phi$. At late times therefore 
it is the state of $\Psi$ 
which fixes the anti-commutator of $\phi$. 
This allows to remove the restriction that initially
the density matrices factorizes. If one is interested by the late
time behavior, only $N$ matters.

In brief, we have recalled two important results.
 First, 
at late time, the Heisenberg field $\phi$ reduces to its
driven term, the second term of eq. (\ref{trues}), 
since both its commutator and anti-commutator
are determined by those of ${\Psi^o_i}$. 
Second, 
 only two (real) kernels 
 determined by the environment
govern the two-point functions of $\phi$, namely
$D$ and $N$ of eqs. (\ref{d}, \ref{n}). Therefore the set
of 
environments (Gaussian or {\it not} Gaussian)
possessing 
the same kernels will give rise to the same 2pt functions. 
Hence they should be viewed as forming an equivalent class. 
The degeneracy can 
be lifted by considering correlations with 
observables containing the operators $\Psi_i$,
or higher order correlations functions of $\phi$ 
(for non-Gaussian environments),
two possibilities we shall not discuss 
in this paper.

 To compute 
   $G_c$ and $G_a$, 
  two different routes 
  can be adopted. 
  When $g_i(p)$ and $\Omega(p)$ are constant, 
  one should work in Fourier transform because 
  the equations can be algebraically solved, in full 
  generality. 
  Instead when $g_i(p)$ and/or $\Omega(p)$
  are time-dependent, as it will be the case 
  in expanding universes and in curved space-times,
 it is appropriate (but not necessary)
 to exploit the above mentioned degeneracy by choosing 
 the $\Psi_i$ and
  their frequencies $\Omega^2_i(p)$ 
  so as to simplify the time dependence of the equations. 
In the text, we proceed with time dependent approach.
In Appendix A, we present
the Fourier analysis which is 
straightforward. 
We invite 
the reader unfamiliar with the Quantum Mechanical 
treatment of dissipation to read it.

 \subsection{
Time dependent settings} \label{TDS}

In Appendix B, we provided a class of 
models characterized by the power of 
$p/\llv$
which specifies how dissipative effects grow, 
see eq. (\ref{iDRn2}).
This class covers the general case and 
can be used as a template to study the 
consequences of dissipative effects. 
In addition, as noticed after eq. (\ref{DSr}), the dissipative effects 
are governed by 
$D = \Sigma_i g^2_i R^o_i$. Therefore
all environments 
delivering the same kernel 
give rise to the same (stationary) phenomenology.

In this Section we exploit 
this freedom 
having in mind 
the transposition of our model from Minkowski space to 
 cosmological, and thus time-dependent, metrics. 
 Therefore the 
 selected 
 models should possess two-point functions with
 simple properties when expressed in the terms of time 
 (as opposed to Fourier components).    
  The core of the problem is that 
  $G_a$ 
 depends, see eq. (\ref{Gags}), on the
 retarded Green function of $\phi$ which is not
 known. 
 Indeed $G_r$ is only implicitly defined as a solution of 
 eq. (\ref{Gr}) which 
 is, in general, a {\it non-local} 
 differential equation.  
 We are thus led to choose $\Psi$ 
 in order for this equation 
 to be local. 
   This implies
 that the retarded Green function of $\Psi$ appearing in 
 eq.  (\ref{Gr}) 
 be proportional to $\delta(t-t')$. (Since 
 this requirement concerns the Green function of the environment
 it does not restrict the phenomenology of $\phi$.)

 Given this aim, we select the models 
  defined by the action
 \ba
   S_T^{(n)}({\bf p}) &=& {1 \over 2 } \int\!dt\ \phi_{\bf p}^* 
 (- \partial_t^2 -  \om_p^2 ) \phi_{\bf p} 
+ {1 \over 2 }  \int\!dt\int^\infty_{-\infty}\!dk\ \Psi^* ({\bf p},k) 
 ( - \partial_t^2 - \left({\pi\llv k}\right)^2 
 ) \Psi({\bf p}, k)
\nonumber\\
&& + {g \llv} \int\! dt \int^\infty_{-\infty}\! dk  \, 
\left({p \over \llv}\right)^{n+1} 
\phi_{\bf p} \, 
\partial_t \Psi^* ({\bf p},k) \, .
 \label{dynM}
 \ea
 In $S_{\phi \Psi}$ have factorized out a factor of $\llv$ so that the 
coupling constant $g$ is dimensionless. 

When compared with the action of
eq. (\ref{Sd}), we have replaced the discrete index $i$ 
by the integral over the dimensionless variable $k$. 
As recalled in Appendix A, 
the spectrum of the environment must be continuous 
to have proper dissipation, see discussion after eq. (\ref{DSr}).
The variable $k$ can be viewed 
as a momentum, in the units of $\llv$, in a flat extra spatial dimension. 
The relationship with some Brane World Scenarios in then clear~\cite{Libanov:2005yf,Libanov:2005nv}.
In the 'atomic' version of 
this model~\cite{Raine:1991kc,Massar:1900vg}
 which has inspired us, 
the radiation field $\Psi$ is a massless 2 dimensional field 
propagating 
in the dimension associated with $k$.

We have also 
introduced an additional time derivative 
acting $\Psi$ in $S_{\phi, {\bf \Psi}}$.
This choice leads to the above mentioned $\delta(t-t')$.
Indeed, on the one hand, taking account this extra derivative, the continuous
character of $k$, and the fact that $g$ is independent of $k$,
 eq. (\ref{eqtosimpl}) becomes
 \be
 ( \partial_t^2 + \om_p^2 ) \phi_{\bf p} = 
   g_n\, 
  \partial_t  \int 
  \! dk   \Psi^o({\bf p},k, t)
   -  \, g_n  \,  \partial_t  \int dt' 
   \int\! dk   R^o(t,t'; k, p) \, 
   \partial_{t'} \big( g_n \phi_{\bf p}(t')\big) \, ,
 \label{eqsimpl}
 \ee
 where $g_n \equiv g \llv \left({p / \llv}\right)^{n+1} $.
On the other hand, for each 3-momentum ${\bf p}$, 
${\bf \Psi}= \int dk \Psi(k)$ 
 is 
 a massless 2-dimensional free field. 
In Fourier components, its retarded Green function 
is given by $R^o(\om, k)= 1/(-(\om+ i \e)^2 + ({\pi\llv k } )^2)$. 
Hence ${\bf R}^o(t,t')$ obeys
 \be
\partial_t {\bf R}^o(t,t') \equiv
\partial_t
\int^\infty_{-\infty} {d\om \over 2 \pi}
 \int^\infty_{-\infty} \! dk \, R^o(\om, k)\, 
 e^{-i \om(t-t')} 
= 
{\delta(t-t')
\over \llv}, 
\label{deltat} 
\ee
which is the required property to simplify eq. (\ref{eqsimpl}).

When $g$ is constant,
the retarded Green function of $\phi$ associated to eq. (\ref{eqsimpl})
obeys the following {\it local} equation 
\be
 [    \partial_t^2 +   {g_n^2 \over \llv} 
  \partial_t + \om_p^2     ] \, G_r(t,t', p) = \delta(t-t')\, ,
 \ee
 To make contact with Appendix A and B, let us rewrite this equation
 in Fourier transform: 
 \be
 [  - \om^2 - ig^2   {\om \over \llv}\,  p^2 \big( { p \over \llv}\big)^{2n}
  + \om_p^2     ] \, G_r(\om, p) = 1 \, .
 \ee
 We thus see that Re$\Sigma_r=0$ and that Im$\Sigma_r$ is 
 (exactly) given by $g^2 $ times that of  eq. (\ref{iDRn2}).
Thus, even though we have chosen a simple form for ${\bf R}^o(t,t')$, 
the above action 
delivers the $n$ dissipative behaviors of Appendix B
by choosing the appropriate power of $p/\llv$ 
in 
$S_{\phi \Psi}$. 
%
%

  When $g$ and $\om_p^2$ are arbitrary
 time-dependent functions, the Fourier analysis looses its power.
 However, in time dependent settings, 
 our 
 $G_r$ still obeys a local equation: 
 \be
 \left[    \partial_t^2 + 2\tilde \gamma_n  
  \partial_t + [ \om_p^2 + \partial_t  \tilde \gamma_n
  ] \right] 
  G_r(t,t') = \delta(t-t') \, .
 \label{tdefGr}
 \ee
 where the $n$-th decay rate $\tilde \gamma_n(t) = g^2(t) \gamma_n$ 
 is now a definite time dependent function. 

 \subsection{Covariant description} 
 
 We now provide the covariantized expression of the action
 $S_T = \int d^3p\, S_T({\bf p})$, where $S_T({\bf p})$
 is given in eq. (\ref{dynM}).
This expression will then be 
used 
to define our theory 
 in curved backgrounds.\footnote{\label{f2}This procedure perhaps 
 requires further explanation since 
  $\Psi$ is not a fundamental field but 
  "nothing more than a convenient
 parameterization of some environment degrees of freedom." 
First, $\Psi$ has not been introduced to parameterize 
  the dissipative effects arising from any theory, but only those
  from theories obeying the Equivalence Principle, 
  see \cite{Libanov:2005yf} for a prototype. Second, 
from the point of view of cond-mat physics, 
it could a priori 
seem inappropriate to proceed to a covariantization, since 
 in most 
  situations (heat bath) 
  there is a preferred frame which is globally defined. 
 However, 
 when the system is non-homogeneous (e.g. a fluid
  characterized by a non-homogeneous flow), 
 low energy fluctuations effectively live in a 
 curved geometry\cite{Unruh:1980cg}.
 Moreover, in this case, 
 short distance effects, i.e. dispersive (or dissipative) 
 effects, are 
 covariantly described 
 when using this metric
 because they arise {\it locally}~\cite{Unruh:1994je,Jacobson:1996zs}.  
 Hence 
 the covariantization is both a necessary step 
 to implement the EP in our settings and a 
 property that emerges in cond-mat physics.}
 
 Two steps should be done. We need to go from $\bf p$ considerations 
 to a local description, and express the various actions in terms of 
 the unit vector field $l^\mu$ and the spatial metric $\perp^{\mu \nu}$. 
 Both are straightforward and, in arbitrary coordinates, the action reads 
 \ba
   S_T&=& - {1 \over 2 } \int\!d^4x \sqrt{-g}\,  g^{\mu \nu}
   \partial_\mu \phi \partial_\nu \phi
\nonumber\\
&&\,  + {1 \over 2 } \int\!d^4x \sqrt{-g} 
 \int\!dk\
 \Big[ \big(  l^\mu l^\nu - 
 c^2_\Psi \!\perp^{\mu \nu} \big) 
\partial_\mu  \Psi(k)  
 \partial_\nu\Psi(k)
  - \left({\pi\llv k}\right)^2 
  \Psi^2(k) \big) \Big]
\nonumber\\
&& +   { g \llv}\int\!d^4x \sqrt{-g}
\Big( \big({ \Delta 
\over \llv^2 }\big)^{(n+1)/2} 
\phi \Big)  \, 
l^\mu \partial_\mu  \int\! dk \Psi (k) \, ,
 \label{CdynM}
 \ea
where 
the symbol $\Delta$ is the Laplacian on the three surfaces
orthogonal to $l^\mu$.
 
 We have slightly generalized the action of eq. (\ref{dynM})
 by subtracting 
  $c^2_\Psi\!\perp^{\mu \nu} $ to the kinetic term of $\Psi$,
 where $c^2_\Psi \ll 1$.
 With this new term, 
 the $\Psi(k)$ are now massive fields which propagate 
 with a velocity whose square is bounded by $c^2_\Psi$.
 In addition they now possess a well defined energy-momentum
 tensor which can be obtained by varying their action with respect to
 $g^{\mu \nu}$. To obtain 
 the simplified expressions 
 we have used (and shall still use), 
 the (regular) limit $c^2_\Psi \to 0$ should be taken.

 In this limit, the (free) retarded Green function 
 of the ${\bf \Psi}$ field 
 obeys a 
 particularly simple equation 
 when expressed in space-time coordinates: 
 \be
 l^\mu {\partial \over \partial x^\mu} \int dk R^o(x,y;k) = 
 l^\mu {\partial \over \partial x^\mu} {\bf  R}^o(x,y) ={ 
 1\over \llv}
  {\delta^4(x^\mu -y^\mu)\over \sqrt{-g}}\, .
  \label{covdelta}
 \ee
 On the r.h.s, one finds 
 the 
 delta function
 with respect to the invariant measure $d^4x \sqrt{-g}$. 
 This equation is nothing by the covariantized and "localized" version
 of eq. (\ref{deltat}).
 Its physical meaning is clear. 
In our model, the back-reaction
 of $\phi(x)$ onto itself through ${\bf \Psi}$ is local.
 
In spite of this, the equations of motions 
 do not have a particularly simple form.  
 Using the condensed notion ${\bf \Psi} = \int dk \Psi$, one gets
 \be
 { 1 \over \sqrt{-g}}\, \partial_\mu \sqrt{-g} \, g^{\mu \nu} \partial_\nu \, 
\phi(x) =  g\llv {  1\over \sqrt{-g}} 
 \big({ \Delta \over \llv^2 }\big)^{n+1 \over 2} \sqrt{-g}\, 
  l^\mu\partial_\mu {\bf \Psi}(x) \, ,
\label{coveomphi}
\ee
where the interacting ${\bf \Psi}(x)$ field is 
\be
{\bf \Psi}(x)= {\bf \Psi}^o(x) 
-  \, g\llv 
\int\! d^4y \sqrt{-g}\,  {\bf R}^o(x,y) \Big[ { 1 \over \sqrt{-g}}\, 
{\partial_\mu }
\Big(l^\mu   \sqrt{-g}  \big({ \Delta \over \llv^2 }\big)^{n+1 \over 2} \phi(y)
\Big)\Big] \, .
\label{covpsisol}
\ee
 When inserting eq. (\ref{covpsisol}) in eq. (\ref{coveomphi}), 
 using eq. (\ref{covdelta}), one verifies that the dissipative term
is local and first order in $l^\mu \partial_\mu$. Therefore, as 
expected, dissipation occurs along the preferred direction specified
by the vector field $l^\mu$.

\subsection{Dissipative effects in curved background geometries}
\label{curvedbg}
%
%
 
 To define a dissipative QFT in an arbitrary curved geometry, 
 one needs some principles. 
 From a physical point of view, we adopt the Equivalence Principle,
 or better what can be considered as its generalization in the presence
 of the 
 unit time-like vector field $l^\mu$. We are in fact dealing
 with two (set of) dynamical fields, the $\phi$ field we probe,
 and the ${\bf \Psi}$ field we do not;  
 but also with two background fields $g_{\mu \nu}$ and $l^\mu$.  
 The Generalized Equivalence Principle means that the action densities of the 
 dynamical fields be given by scalar functions (under general coordinate 
 transformations) which coincide
 to those one had in Minkowski space time for a homogeneous and static
 $l^\mu$ field, i.e. those of eq. (\ref{CdynM}).
 
 However, the densities are 
 not completely fixed by the GEP. In this we recover what was 
 obtained with the EP: For 
 a scalar field in a curved geometry, there was always the 
 possibility of considering a non-minimal coupling to gravity
 by 
 adding to the Lagrangian 
 a term proportional to $R \phi^2$. 
  In the present case, the ambiguity is larger because
   $l^\mu$ defines 
 new scalars, the first of which is the expansion 
 $\Theta = \nabla_\mu l^\mu$, 
 where 
 $\nabla_\mu $ is
 the covariant derivative with respect to $g_{\mu\nu}$.
 The ambiguity can only be resolved by 
adopting some additional principle, such as 
the principle of minimal couplings which forbids 
adding densities containing these scalars. 

Rather than adopting it, we shall choose the non-minimal coupling 
so as to keep eq. (\ref{covdelta}), i.e. so that the 
{\it locality} of the back-reaction effects of $\phi$ through $\bf \Psi$
be preserved. 
This choice maintains the simplicity of the equations
of motion in curved backgrounds,
but is by no means necessary.
Starting from eq. (\ref{CdynM}), 
the locality is preserved by replacing 
 in $S_\Psi$ and $S_{\Psi \phi}$ 
 \ba
l^\mu \partial_\mu\Psi_k \to 
{\cal D}_l \Psi_k \equiv 
l^\mu \partial_\mu\Psi_k + {\Psi_k \over 2} \  \Theta 
= {1 \over 2}   \left(  l^\mu
\nabla_\mu \Psi_k +  \nabla_\mu \, [l^\mu \Psi_k] \right)\, .
\label{covl}
 \ea

To simplify the forthcoming equations, we use the fact
 that one can always work in "preferred" coordinate systems 
 in which the shift $l^i$ vanishes and in which 
 the preferred time is such that $l_0 = 1$. 
 (We assume that the set of orbits of $l^\mu$ is complete 
 and without caustic. In this case, every point of the manifold
 is reached by one orbit.)
In these coordinate systems, it is useful to work 
with rescaled fields $\Psi^r \equiv (-g)^{1/4}\,
 \Psi$ because the above equation simplifies 
\be
{\cal D}_l \Psi_k
= (-g)^{-1/4} \, l^\mu  
\partial_\mu ( (-g)^{ 1/4}\,\Psi_k)= (-g)^{-1/4} \, l^\mu  
\partial_\mu \Psi_{k}^r \, ,
\label{covl2}
 \ee
 since
$ \Theta = (-g)^{-1/2}\partial_\mu[( -g  )^{1/2}l^\mu]$.
%
%
 Notice also that there exists a 
 subclass of 
 background fields $(g,l)$, for which one can find 
  coordinate systems such that
 {\it both} the shift $l^i$ and $g^{oi}$ vanish.
 In these {\it comoving} coordinate systems, 
 the above equations further simplify  since only the spatial part of the 
 metric matters because
 $ -g=  h_c$ where 
 $h \equiv {\rm det}(\perp_{ij})$.
  ~\footnote{\label{intfnote} 
It is clear that this is the case when $l^\mu$ coincides with 
the cosmological frame 
and when one uses comoving coordinates  
$ds^2 = -dt^2 + a^2 dx^2$ since $l^\mu=(1,0)$. 
However, in certain cases one should search for the
"comoving" coordinate system.
To illustrate this point, 
 consider the former situation 
 in Lema\^itre coordinates
 $X=a \, x$. In this case one has $ds^2 = -dt^2 +(dX - V dt)^2$, 
 where the velocity is 
 $V=HX$. 
 The spatial sections are now the Euclidean space with $h=1$, and 
 the (contravariant) components of the unit vector field are $l^\mu= 1, V$. 
 To compute $h_c$ one should solve the equation of motion of comoving 
 (free falling) 
 observers $dX -V dt = 0$, and use the initial position
 as new coordinates. 
 This procedure is 
explicitely 
done in \cite{Parentani:2007uq} when 
starting with Painlev\'e-Gullstrand 
coordinates to describe the
black hole metric and using a freely falling frame.} 

Having chosen this non-minimal coupling, 
one verifies that the kinetic term 
of the rescaled fields $\Psi_r$ 
is insensitive to the "curvature" of both $g_{\mu \nu}$
and $l^\mu$ (when the limit $c^2_\Psi \to 0$ is taken).
Moreover the differential operator which acts on the
retarded Green function of $\Psi_r$ in the equation 
of motion of $\phi$, see eqs. (\ref{coveomphi}, \ref{covpsisol}), is also "flat"
thereby guaranteeing that the modified version
eq. (\ref{covdelta}) still applies, that is
%
%
\ba
 {\cal D}_l \, 
{\bf  R}^o(x,y)&=&    (-g(x))^{-1/4}   \, 
 l^\mu {\partial
\over \partial{x^\mu}} \Big({\bf  R}_r^o(x,y)\Big) \,  (-g(y))^{-1/4}
  \nonumber \\
&=& {1 \over \llv} { \delta^4(x^\mu -y^\mu)
  \over \sqrt{-g}}\, ,
 \label{covcurveddelta}
 \ea
where ${\bf  R}_r^o(x,y)$
is the retarded Green function of the rescaled field $\Psi_r$.
It obeys (in preferred coordinate systems) 
$\partial_t {\bf  R}_r^o = \delta^4/ \llv$,
and "defines" the retarded Green function 
${\bf  R}^o= (-g)^{1/4}\,{\bf  R}_r^o \, (-g)^{1/4}$ which is a bi-scalar.
Hence the equations of motion in an arbitrary background "tensor-vector metric" specified by the couple ($g_{\mu\nu}, l^{\mu}$), are given by eqs. 
(\ref{coveomphi}, \ref{covpsisol}) 
with the substitution of eq. (\ref{covl}).

Several remarks should be made.
First, from the simplified 
equation $\partial_t {\bf  R}_r^o =\delta^4/ \llv$
it might seem that the background tensor metric $g_{\mu \nu}$ plays no role.
This is not true, since 
it is used to normalize the field $l^\mu$ at every point. 

Second, in the limit $c^2_\Psi \to 0$,
the (rescaled) $\Psi_k$ fields define a new kind of field. 
They propagate
in an 
effective space-time given by the time development
of the 3-dimensional set of orbits of the $l^\mu$ field. 
Indeed, at fixed $k$, $\Psi_k(x)$ can be decomposed in 
non-interacting local field-oscillators, each of them evolving separately
along its orbit. This situation is similar to the long wave length (gradient-free)  
expression of 
\cite{Salopek:1992kk}.
  In the absence 
of $l^\mu$, the 
geometry must be (nearly) homogenous for the 
action 
to posses this decomposition. However,  when 
$l^\mu$ is given, 
 one can identify, even in non-homogeneous metrics, 
each space-time point in an invariant way by the spatial position of the
corresponding
"preferred" orbit at some time, and the proper time along the orbit (as long
as $l^\mu$ has no caustic). 
We can thus build covariant actions exploiting this possibility
and consider fields composed of a dense set of
 local oscillators 
at rest with respect to $l^\mu$.
The fields $\Psi_k$ we use belong to this class of fields. 

 \section{Dissipative effects in cosmology }
 
 \subsection{The action}
 
 Our aim is to describe dissipative effects 
 in an expanding homogeneous universe 
 when the vector field $l^\mu$ is aligned along 
 the cosmological frame, when the dissipative effects are 
 known 
 in Minkowski space, and when implementing the Equivalence Principle.

 In this case, to get the action 
 we simply 
 consider eq. (\ref{CdynM}) (with the curved metric modifications
 discussed in the former subsection) in a FLRW metric.
 Using comoving coordinates,
  \be
ds^2 = - dt^2 + a^2(t) \, d{\bf x}^2 
\, ,\label{RW}\ee
the components of $l^\mu$ 
are 
$(1,\bf 0)$.
To simplify the notations, we use
 the conformal time $d\eta = dt/a $ and work with the 
 rescaled fields $\phi_r= a \phi$ and
$\Psi_r = h_c^{1/2}\Psi = a^{3/2}\Psi$. 
 Dropping these $r$ indices, working in Fourier transform 
 with respect to 
  $\bf x$, 
the  non-minimally coupled action associated with the replacement
of eq. (\ref{covl})  is 
 \ba
   S_T^{(n)}({\bf p}) &=& {1 \over 2 } \int\! d\eta \ \phi_{\bf p}^* 
 (- \partial_\eta^2 -  \omega^2_p(\eta) )  \phi_{\bf p}
\nonumber\\
&&+ {1 \over 2 }  \int\! dt\int\! dk\ \Psi^* ({\bf p},k) 
 ( - \partial_t^2 - \left({\pi\llv k }\right)^2 
 ) \, \Psi({\bf p}, k)
\nonumber\\
&&+ \int\! d\eta  \, g_n(\eta)\,  
\phi_{\bf p} \, 
\partial_\eta \int\! dk \Psi^*({\bf p},k) \, .
\label{dynRW}
 \ea
The conformal frequency
  $\omega^2_p(\eta)= p^2 - 
 \partial^2_\eta a/ a$ is that 
 of a rescaled 
 minimal coupled massless field. In this expression, as everywhere in this Section,
  $p$ is now the conformal (dimensionless and constant) wave vector. 
 The time dependent coupling coefficient 
 is \be
 g_n \equiv g\  a^{1/2}
  \ \llv \left({p / a \llv}\right)^{n+1} \, .
  \ee 
  Its dependence in $a$ (given that of $p$)  
  follows from 
 having implemented the Generalized  Equivalence Principle (GEP)
  which determines the powers of $a$ for each term in the action.
 As we shall see, 
 this relative power   
  guarantees that all $\phi_p$ will be damped
  at a {\it fixed and common} proper scale (in the adiabatic approximation).  
  Had we started with an action of the type (\ref{dynRW}) without relying on the
  GEP, the dependence in $a$ would have been
  arbitrary, and the proper scale at which modes would have been damped
  would have run as well. 
%
  

In the same vein, one sees 
that the proper frequency of the $\Psi$ fields stays constant.
This follows from the GEP 
but also 
from our choice of non-minimal couplings. 
 In this paper, we want 
 indeed to analyse the phenomenology of 
  dissipative effects when the 
 new degrees of freedom are and stay in their ground state. 
 One could have chosen  
  more complicated models in which $\Psi$ is 
  parametrically excited. This would be the case when 
  introducing a non zero velocity $c_\Psi$, and 
  with minimally coupling 
  (by dropping $\Theta$ 
  on the rhs of $\equiv$ in  eq. (\ref{covl})).
However, 
when  $\Psi$ is taken  massive (i.e. by restricting
  the $k$ range to $\vert k\vert  > 1 $),
  or discretizing $k$ as it would be the case for Kaluza-Klein modes,
  and if $\llv \gg H $, the phenomenology of all these models coincide 
  since the amplitude for parametric excitations will be exponentially damped.
  In this regime there is thus no gain in studying more complicated actions
  than that given in eq. (\ref{dynRW}).
  

  \subsection{Equation of motion}

  Using the results of the subsection \ref{TDS}, 
  the equation of motion of Heisenberg operator $\phi_{\bf p}$ is 
  \be
  \big(\partial_\eta^2 + 
  2  \gamma_n\,  \partial_\eta  + ( \omega^2_p(\eta) 
  + \partial_\eta\gamma_n)
  \big)  \, \phi_{\bf p} =  
  g_n \, \partial_\eta {\bf \Psi}^o({\bf p})\, ,
  \label{cosmoeom} 
  \ee
  where 
  the decay rate in conformal time is, 
  \be
  \gamma_n(\eta) = { g^2_n \over 2 \llv}
  = {1 \over 2} ( a\llv ) \big({p \over a \llv}\big)^{2n+2}\, .
  \ee
  It is dimensionless, as it should be. It should be compared with the
  comoving frequency $p$ to get the relative strength of dissipation. 
  One obtains
  \be
   {\gamma_n(\eta)\over p} = {1 \over 2} \big({p \over a \llv}\big)^{2n+1} = 
   {1 \over 2} \big({p_{phys} \over \llv}\big)^{2n+1}\, . 
   \label{gammaninfl}
  \ee 
  In the last equality we have re-introduced 
  the proper momentum $p_{phys}= p/a $.
  With this equation we verify that, at any time in an expanding universe
  and for every mode $\phi_p$,
  the relative strength of dissipation is simply that of Minkowski space 
  evaluated at the corresponding energy scale, see 
   eq. (\ref{decayn}).
  This directly results from 
  having implemented the GEP. 
  
  Notice however that the equation of motion in an expanding
  universe contains a frequency shift
  \be
   { g^2_n \over \llv}\partial_\eta (\ln g_n  )= p^2  \, \big({aH \over p}\big)\, 
   \big({p \over a \llv}\big)^{2n+1} \, .
   \label{levels} \ee 
  We have factorized out the unperturbed frequency square to get the
  relative value of the shift. It vanishes both when dissipation is negligible
  and when $aH/ p \ll 1$, i.e. when the expansion rate $H$ is negligible
  with respect to the proper momentum. From this expression
  we can already conclude that it cannot play any role
  when the two relevant scales $H$ and $\llv$ are well separated, 
  i.e. when 
  \be
  \sigma \equiv { H \over \llv }  \ll 1 \, .
    \label{psigma}
    \ee
   Indeed when the physical momentum is high and of the order of $\llv$, 
   the relative frequency shift proportional to $\sigma$, and
   when the physical momentum is of the order of $H$ (at horizon exit, 
   see Figure 1), it is  proportional to $\sigma^{2n+1}$. 
     
  In quantum settings however, it is not sufficient that the equation
  of motion possesses its Minkowskian form because the 
   state of the system, and therefore the 
   observables,
   might be affected by the combined effect
  of dissipative effects and the expansion rate. 
  
 \subsection{Power Spectrum} \label{Power Spectrum}
  
  Let us consider 
  the power spectrum of a scalar massless test field 
  (which is the relevant case for
  gravitational waves and density perturbations)
  both in the standard free field settings and
  in the presence of dissipative effects.

%
%
%

\subsubsection{Free settings}

The equation of motion is the same as in eq. (\ref{eqom}) with
\be
\omega_p^2 =p^2 - {2 \over \eta^2} \, .
\label{eomfree}
\ee
For simplicity we consider the inflationary background to be 
de Sitter where $a(\eta) = - 1/H \eta$ and $\eta < 0$.
At the onset of inflation, 
when $p_{phys}/H = p /a_{in}H =  p ( - \eta_{in}) \to  \infty$, 
the positive frequency modes are
\be
\phi^{in}_{\bf p} = {1 \over (2 p)^{1/2}} \left( 1 - {i \over p\eta } \right)\, e^{-i p \eta} \, .
\label{inm}
\ee
When 
 $ p \vert \eta \vert = 1$,  
the physical wave length $\lambda = a/p$ 
becomes larger than the Hubble radius. 
Near that "horizon-exit", 
$\omega_p^2$ 
flips sign, the mode stops oscillating and starts to grow like $a$.
At late time with respect to horizon exit, 
when $\vert p\eta \vert \ll 1$, this implies that 
the power spectrum 
 becomes a constant, as we now recall.

When inflation lasts long enough, i.e. when  the number of extra e-folds
obeys $N_{extra} \gg 1$, see eq. (\ref{Nex}), all observable modes $\phi_{\bf p}$ are in their ground state 
at the onset of inflation (simply because this is the only state compatible 
with inflation~\cite{Parentani:2004ta}). 
In this case, 
the anti-commutator of the free field, see eq. (\ref{Gags0}), 
when evaluated at equal time $\eta$ is simply
given by
\be
  G^{free}_a(\eta,p)
  =  \vert \phi^{in}_{\bf p} \vert^2 = {1 \over (2 p)} 
 \left( 1 + {1 \over (p\eta)^2 } \right)\, .
 \label{invac}
\ee
Then the power spectrum of the 
physical (un-rescaled) field 
given by 
\be
P^{free}_p(\eta) =  {p^3\over  2 \pi^2} {G^{free}_a(\eta,p) \over a^2(\eta)} 
= \big({ H_p^2 \over 2 \pi}\big)^2 \, (1 + (p\eta)^2) \, ,
 \ee
 becomes constant after horizon exit.
We have added a $p$ subscript to $H$ because in slow roll inflation,
the relevant value of $H$ for the $p$-mode is that evaluated
at horizon exit, i.e. $H_p = H(t_p)$, where $t_p$ is given by
$p/a = H$. The above equation shows that $P_p$ acquires some 
scale dependence only through $H_p$. Similarly the
{\it deviations} from this standard behavior stemming from some
UV modification of the theory will also 
depend on $p$ through $H_p$ (and its derivatives).
For explicit examples, we
 refer to \cite{Niemeyer:2002kh} 
 where the modifications of the spectrum 
stem from the fact that $p \eta_{in}$ is taken large but finite,
and also to \cite{Campo:2007ar} wherein the UV cutoff is endowed with a
finite width. This last case bears many similarities
with the dissipative settings we now study.



\subsubsection{Dissipative settings}

In the presence of dissipation, the expression for 
$G_a$ 
radically differs from the above. 

At the level of the Heisenberg operator, 
when dissipative effects grow with
the energy (as we suppose it is the case), 
the decaying solution of eq. (\ref{cosmoeom})
is completely erased (unless one fine-tunes $N_{extra}$, see eq. (\ref{Nex}),
so as to keep a residual amplitude). That is, in inflation
 the mode operator is entirely
given by its driven term, the second term in eq. (\ref{trues}). 
Then the power spectrum is also 
purely driven 
and given 
by the second term of eq. (\ref{Gags}):
\be
G_a^{driven}(\eta,p) = \int\! d\eta_1 \int \!d\eta_2 \,
G_r(\eta, \eta_1, p)\,  G_r(\eta, \eta_2, p) \, {\bf N}(\eta_1, \eta_2, p)\, ,
\label{Gadriven}
\ee
where $G_r$ 
is the retarded Green function,
solution of eq. (\ref{cosmoeom})
with $\delta(\eta - \eta_1)$ as a source, 
and where the kernel ${\bf N}$ is the anti-commutator 
of $g_n \, l^\mu\partial_\mu {\bf \Psi}_p$, the source of eq. (\ref{cosmoeom}).

Eq. (\ref{Gadriven})
tells us that only the 
state of the 
environment matters. In other words, 
because of the strong dissipation at early times,
the power spectrum is independent of
the initial state of $\phi$, what ever it was. 
In spite of this, 
 when eq. (\ref{psigma}) is satisfied, 
and when the environment is in its ground state, the predictions
are unchanged, i.e. the power spectrum obtained from 
$G_a^{driven}$ coincides with that obtained with $G_a^{free}$.
To show this let us study $G_r$ 
and 
${\bf N}$.

\subsubsection{Dissipative Green functions and noise kernel}

The retarded Green function is of the form
\be
G_r(\eta, \eta_0, p) = \theta(\eta - \eta_0) \, 
e^{- \int^{\eta}_{\eta_0} d\eta' \gamma(\eta')}
\times 2{\rm Im} \Big( \tilde \phi_p^*(\eta) \,  \tilde \phi_p(\eta_0) \Big)\, ,
\label{Grinfl}
\ee
where (in the under-damped regime) the modes $\tilde \phi_p$ are 
unit Wronskian positive frequency solutions 
of eq. (\ref{eomfree}) with a frequency square 
given by
\be
(\om^{ef\!f}_p)^2 =\om^{2}_p - \gamma_n^2 =  p^2 \, ( 1 - {2 \over p^2\eta^2} -
{\gamma_n^2 \over p^2}) \, .
\label{omeff2}\ee
In this, we obtain 
a time-dependent version
of the stationary case, see eq. (\ref{omeff}).
With the second expression, we verify that when eq. (\ref{psigma}) is 
satisfied, the two corrections terms are never simultaneously relevant,
which guarantees, as we shall discuss below, that dissipative
effects can be studied in the quasi-stationary approximation.
Notice also that the frequency shift $\partial_\eta \gamma$ present is 
eq. (\ref{levels}) drops out from
$\om^{ef\!f}$.

Eq. (\ref{Grinfl}) implies 
that
\be
G_r(\eta, \eta_0, p) \to 0 \quad {\rm when} \quad 
\eta_0 \to
\eta^\Lambda_p \, ,
\label{erase}\ee
where 
$\eta^\Lambda_p$
is the "$\llv$-exit" time of the $p$-mode defined by
$p/a(\eta^\Lambda_p) = \llv$, or $\gamma/p = 1/2$, see eq. (\ref{gammaninfl})
and the Figure. \begin{center}
\includegraphics[width=7cm]{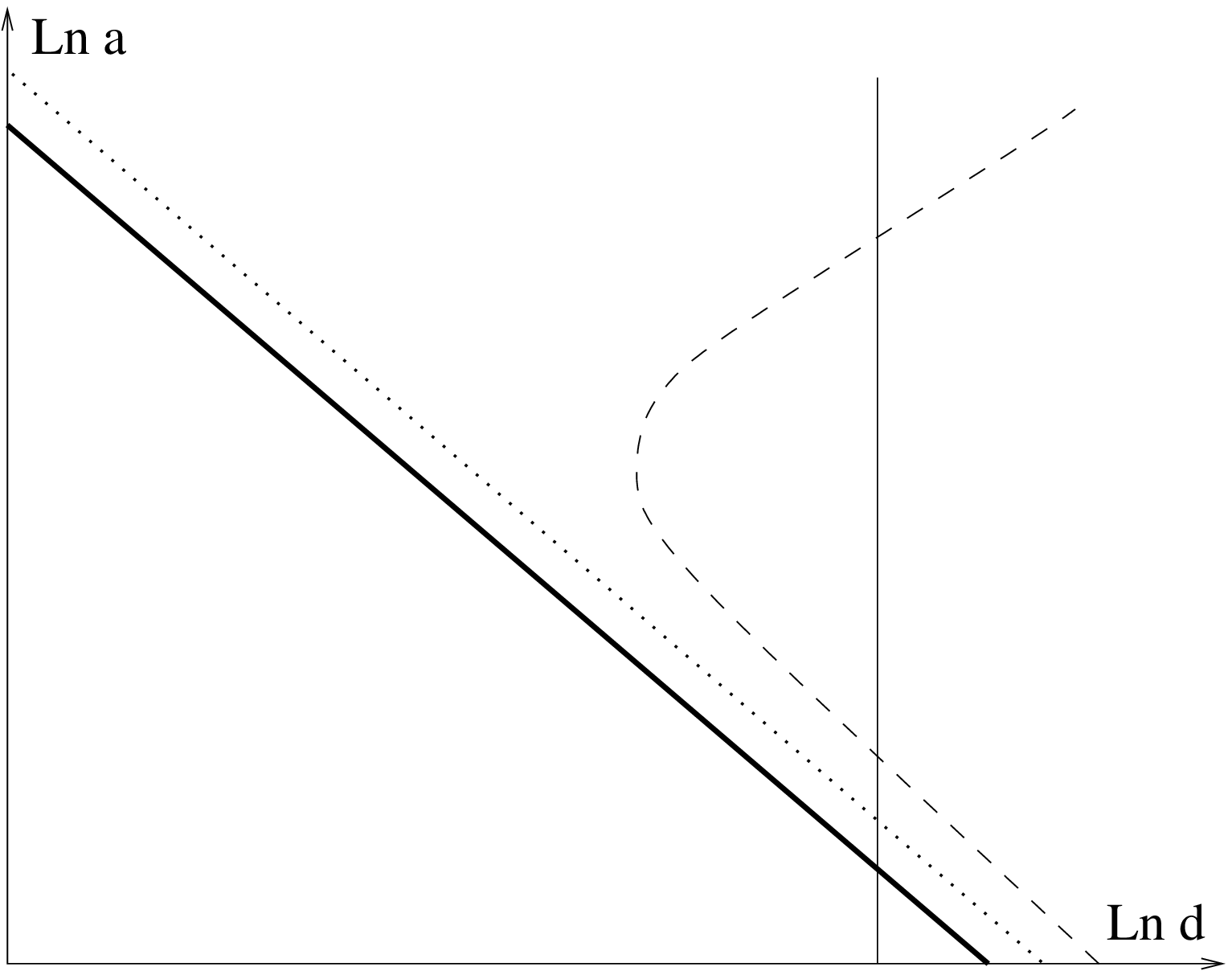}
\end{center}Figure caption.
{\it 
We have represented in a log-log plot and by a dashed line 
the evolution of $d_H = R_H/a$, the comoving Hubble radius, 
as a function of $a$,
 both during inflation $(d_H \propto 1/ a)$ and during the
radiation era $(d_H \propto a)$. We have represented by a thick line
the trajectory of the cutoff length 
 $d_\Lambda = 1/a\llv$ of a fixed proper scale
which obeys $1/\llv \ll R_H$ during inflation. 
The dotted line corresponds to an intermediate 
fixed proper length $\lambda$ which obeys
$1/\llv \!\ll\!\lambda \!\ll\! R_H^{infl.}$.
The vertical line represents a fixed comoving scale $d_p = 1/p$. Below the 
cutoff length, all modes are over-damped. When a mode exits the cutoff length,
it becomes under-damped and starts propagating. When it crosses the intermediate 
length $\lambda$, it behaves as a free mode, and it gets amplified 
only when exiting the Hubble radius.
Adiabaticity, which is guaranteed by $1/\llv \ll R_H$, 
guarantees in turn that, near $\lambda$, modes are all born in the Bunch-Davies vacuum 
when the environment $\bf \Psi$ is 
in its ground state.}

Eq. (\ref{erase}) guarantees that no quantum 
coherence is left between late time $\eta$, e.g. horizon-exit,
and what happened for times earlier than $\llv$-exit. 
The physical implication of this is that, what ever transitions happened,
what ever was the quantum state, 
no record is kept at late time.\footnote{
This behavior 
has been described, drawn, 
and sometimes named {\it anti-dissipation}, see 
\cite{Brout:1995wp,Jacobson:1996zs,Niemeyer:2000eh,PhysRep,River2000,BFP2,Parentani:2007mb}.
In fact when tracing backwards in time 
 vacuum configurations 
from the freely propagating regime down to the
dissipative one, dissipation arises towards the past. 
When viewed as a function of $\eta_0$,
$G_r(\eta,\eta_0)$ indeed {\it increases} to the future, thereby showing anti-dissipation.
From eq. (\ref{Grinfl}) we understand the origin of this quantum 
property (without classical counterpart). 
As a function 
of $\eta_0$, first, $G_r$ obeys eq. (\ref{cosmoeom}) with the "wrong" sign 
for the dissipative term, and second, the boundary condition 
is a final one and not an initial one, 
thereby explaining why when $\eta_0 \to \eta$,
the value of the current is "adjusted" to become exactly one.}
Having understood the structure of the retarded Green function, 
let us now briefly consider the commutator $G_c(\eta, \eta_0)$. 
It is given by 
\ba
G_c(\eta, \eta_0, p) &=& -i \Big( G_r(\eta, \eta_0, p) - G_r(\eta_0, \eta, p)
 \Big)\, ,
\nonumber\\
&=& e^{-\vert  \int^{\eta}_{\eta_0} d\eta' \gamma(\eta')\vert }
\times 2i{\rm Im} 
\Big( \tilde \phi_p(\eta) \,  \tilde \phi_p^*(\eta_0) \Big)\, .
\label{Gcinfl}
\ea
It possesses two noteworthy properties.
First, because of the absolute value in the damping exponential term,
it obeys two different local differential equations
depending of the sign of $\eta - \eta_0$. When $\eta > \eta_0$, 
$G_c$ is an homogeneous solution of eq. (\ref{cosmoeom}), whereas 
when $\eta < \eta_0$, it obeys the equation wherein the sign
of the dissipative term has been changed. These two differential
equations can be grouped into a single non-local equation.
Second, because of this "quasi"-local properties, $G_c$ is independent
of $\eta_{in}$, the moment when $\phi$ was put in contact with $\bf \Psi$.

These properties of $G_c$ follow from the fact that its source term,
the $D$ kernel of eq. (\ref{Gcgs}), is "ultra" local
for the $\bf \Psi$ field of eq. (\ref{dynRW}): 
\be
{\bf D}(\eta_1, \eta_2) \equiv [ g_n \, \partial_{\eta_1} {\bf \Psi}(\eta_1)\, ,\, 
  g_n \, \partial_{\eta_2} {\bf \Psi}(\eta_2)\, ]
= g_n(\eta_1) \,  { i \partial_{\eta_2} 
\delta(\eta_1 - \eta_2) \over \llv }\,  g_n(\eta_2)\, .
\ee 
This should be contrasted with the kernel ${\bf N}$, the anti-commutator,
which is non local. 
Indeed, in the vacuum, it is given  
\be
{\bf N}(\eta_1, \eta_2) = g_n(\eta_1) \,  
{ - a(\eta_1) a(\eta_2)
\over \pi \llv (t_1 - t_2)^2 }\,  g_n(\eta_2)\, ,
\ee 
where $1/(t_1 - t_2)^2$ should be understood as the 
derivative of the principal part of $1/(t_1 - t_2)$. 
One gets a simple expression
in terms of the proper time $t$ because 
the proper frequency of $\Psi_k$ is constant in our model (\ref{dynRW}).
We also remind the reader that it is only in the high temperature limit that $N$
would become
proportional to $ \delta(t_1 - t_2)$.

In this paper we shall compute the power spectrum when the $\Psi_k$ are all in their ground state.
The reason for this choice is as follows: As pointed out in the footnote \ref{f2}, the $\Psi_k$ fields 
should be conceived as parametrizing some degrees of freedom in a fundamental theory
obeying the Equivalence Principle. In these theories, when inflation lasts long enough, 
all relevant degrees will be in their ground state, as is the case when dealing with free modes, see
eq. (\ref{invac}). We see no reason why dissipation could possibly invalidate this conclusion.

When the $\Psi_k$ are all in their ground state, the kernel 
${\bf N}$ is non-local.
Therefore 
the anti-commutator $G_a$ 
obeys a non-local equation: eq.
(\ref{cosmoeom}) with ${\bf N}$ as the source.
The physical meaning is that the environment is driving 
$\phi_p$ in an non-instantaneous way (this is unavoidable
in the vacuum since only one sign of the frequency is present). 
The consequence is that the value of power spectrum depends
(to a certain extend) on the 
history of the 
combined evolution of $\phi + \bf \Psi$.
The mathematical implication is that 
an exact calculation of $G_a$ is effectively impossible.

Nevertheless when $H/\llv \ll 1$, these non local effects give only but
sub-leading (non-adiabatic) corrections, 
because the 
evolution of $\phi + \bf \Psi$
consists in a parametric (adiabatic) succession of 
stationary states ordered by
the scale factor $a$.  

\subsubsection{Scale separation and adiabatic evolution}

Given the importance of this result, we shall spend some 
time to explain its root and its validity.
We proceed in two steps. We first show that adiabaticity is sufficient 
for obtaining the standard power spectrum, and then show that 
scale separation, eq. (\ref{psigma}),
is sufficient to guarantee adiabaticity, thereby
generalysing the results of \cite{Niemeyer:2001qe}.

Adiabaticity means that the evolution proceeds slow enough
for inducing no (non-adiabatic) transition (in molecular physics, they
are called Landau-Zener transitions, 
in the present cosmological context they correspond to pair creation
of $\phi$ quanta). 
When this is the case, 
the observables 
take, at any time, the value
they have in the corresponding stationary situation.

As of 
retarded Green function and the commutator,
this principle does not bring any new information because in the WKB
approximation, they are local functions since these 2pt functions 
obey local equations.  
For 
the anti-commutator, adiabaticity guarantees that when $\eta$ and $\eta'$
are close (in the sense that $1 - a(\eta)/a(\eta') \ll 1$), 
its value is well approximated by 
\be
G_a^{driven}(\eta,\eta', p) 
\simeq
 G_a^{statio}(\eta-\eta'; \omega_p(a),g_p(a)) = \int {d\om \over 2 \pi }\, 
e^{i \om (\eta-\eta')}\, 
 G_a(\omega ; \omega_p(a),g_p(a)) \, ,
\label{Gadiab}
\ee
where $ G_a(\om ;\omega_p(a),g_p(a))$ is the Fourier
 component 
calculated 
with the 
values of the frequency $\omega_p(a)$ and the coupling $g_p(a)$
both evaluated with $a= a(\eta)$.
In Appendix A, these Fourier components 
have been algebraically solved for all frequencies and all couplings, 
see eq. (\ref{Gaf}). 
Moreover, since by hypothesis, we are in the vacuum, 
eq. (\ref{fd}) also applies. In other words, the value of $G_a$ 
follows from that of  
the commutator $G_c$.
This is sufficient to guarantee that when the mode $\phi_p$ becomes
free, i.e. much after $\llv$-exit but before horizon-exit ($H \ll p/a \ll \llv$),
eq. (\ref{mie}) applies. Thus, {irrespectively} of what was the coupling
with the environment,
\be
G_a^{driven}(\eta,\eta, p) \to G_a^{free}(\eta,\eta, p) = {1 \over 2 \om_p(a)}\, .
\label{dfree}\ee
With this equation we reach our first conclusion:  in the adiabatic 
approximation and when the environment is in its ground state,  
the modes $\phi_p$ are born
in the usual $in$-vacuum (Bunch-Davies)
as they become, one after the other, freely propagating 
after $\llv$-exit, see the Figure.~\footnote{
This behavior is in accord with the procedure 
of \cite{Danielsson:2002kx,Niemeyer:2002kh,Campo:2007ar} wherein the initial state is imposed on modes
when their proper momentum reaches a given value. 
This radically differs with the ``initial slice'' 
approach~\cite{Schalm:2004qk} wherein the state is imposed at the same time 
on all modes, irrespectively of the value of their momentum. 
From what we see, the Equivalence Principle allows 
only the first to be dynamically realized.}

It now behooves to us to show that scale separation guarantees
that eq. (\ref{Gadiab}) provides a valid approximation for 
$G^{driven}_a$ before horizon-exit, i.e. when the second term in the
parenthesis in eq. (\ref{omeff2}) can still be neglected. 
To this end, we provide an upper bound
for the probability amplitude of getting a
non-adiabatic transition in the under-damped regime, i.e. 
$H \ll p/a \leq \llv$. 
 This amplitude is
governed by the relative frequency change
\be
{\partial_\eta \om_p^{ef\!f} \over (\om_p^{ef\!f})^2 } 
= - {\gamma_n^2 \partial_\eta \ln \gamma_n \over (p^2 - \gamma_n^2)^{3/2}}
= (2n + 1) {H \, \gamma_{phys}^2  \over (p_{phys}^2 - \gamma_{phys}^2)^{3/2}} \, ,
\ee
where $p_{phys}= p/a$ and $\gamma_{phys}= \gamma/a$ 
are the physical (proper) momentum and decay rate.
Therefore, going backwards in time from the free regime
up to $p_{phys}^2 \geq 4 \gamma_{phys}^2$ (i.e. $p_{phys} \leq  \llv$), 
the non-adiabatic parameter steadily grows 
but stays bounded by 
\be
{\partial_\eta \om_p^{ef\!f} \over (\om_p^{ef\!f})^2 } 
<  3 (2n + 1)\,  
\sigma \, \Big( {p_{phys} \over \llv}\Big)^{4n +1} < 3 (2n + 1)\,  \sigma \ll 1 \, .
\ee
This guarantees that the amplitude for 
the system to jump out of the ground state
is bounded by $\sigma$ (up to an overall factor which plays no role).
There is no need to study the stability of the ground state 
in the transitory regime from under-damped to the over-damped modes
 because what ever transitions happened 
their residual impact at late time 
would be suppressed by a factor $e^{-\int \gamma dt } 
\simeq \exp(-1/\sigma(2n + 1)) \ll 1$. This completes the proof
that scale separation guarantees adiabaticity, in agreement with
the conclusion reached in \cite{Libanov:2005yf,Libanov:2005nv}.



\subsubsection{Perspectives: Beyond the adiabatic approximation and scale crossing}

It is first interesting to point out 
that dissipative models are more "robust" than 
dispersive models in that the "driven" power spectrum determined 
by eq. (\ref{Gadriven})
is well defined even when $H  > \llv$, contrary to the fact that most
 dispersive models
  make sense only when scale separation ($H \ll \llv$) is realized. Therefore,
with dissipative models, one can study scale crossing, i.e. the
crossing of $H_p$ through $\llv$ from above to below
during slow roll inflation. 

%
The power spectrum has been studied in this regime when using a particular
 Brane World scenario in \cite{Libanov:2005nv}. 
 We agree with most of the conclusions but not
 with that reached after Eq. (35) according to which 
 ``towards the end of inflation,
 perturbations on the brane are the sum of two independent fields''.
 We do not agree 
  because in those settings as well the field operator is purely driven. 
 Hence we do not see how to decompose it as a superposition of two 
 commuting operators.  
 Moreover, with the approximation used in  \cite{Libanov:2005nv}, 
 it seems that the ETC,
  eq. (\ref{etc}), is violated 
  by a amount proportional to the reported change of the power spectrum.

In any case, the calculation of 
the modifications of the power spectrum introduced by dissipation
is difficult 
because the non-local properties of 
 eq. (\ref{Gadriven}) cannot be neglected. We are not aware
 of any analytical treatment, 
  and we are currently studying 
  the modifications using numerical techniques~\cite{AdamekCNP}. 
 
 
{\it Added note.} Since the present paper was submitted, we have 
obtained several results~\cite{AdamekCNP}. 
In particular, the leading modification of the power spectrum with respect to the standard result
in the regime $\sigma = H/\Lambda \ll 1$, 
i.e. the signature of UV dissipation, 
scales with a power of $\sigma$
which is equal to that of $P/\Lambda$ in $\gamma/p$, 
see eq. (\ref{gammaninfl}). In this respect, dissipative models
behave like dispersive models where the leading modification
scales with a power of $\sigma$ which is that of the first non-linear term
in the dispersion relation, i.e. $2n$ using the parameterization of eq. (\ref{disprel}), see ~\cite{Macher:2008yq}.
In the opposite regime,
when $H > \Lambda$, no universal behavior is obtained.

\section{Conclusions}
In this paper we have obtained the following results. 

First, we have provided a class of unitary 
models defined in Minkowski space which are characterized
by the power of $p/\llv$ which weighs the growth
of dissipation in the preferred frame, see eq. (\ref{iDRn2}). Unitarity 
is achieved by introducing a dense set of additional fields $\bf \Psi$
which induce dissipation through interactions with $\phi$.
In Appendices A and B we have given a
thorough analysis of the Green 
functions of these Gaussian models which cover the phenomenology
of UV dissipative effects in Minkowski space at the level of 2pt functions.

Second, amongst the various actions delivering the same stationary phenomenology,
we have selected one 
which gives rise to a local differential equation for the retarded 
Green function of $\phi$, see eqs. (\ref{eqsimpl},\, \ref{deltat}). 
Using the Equivalence Principle, 
we have 
extended this class of models to arbitrary
curved backgrounds in subsection \ref{curvedbg}, 
thereby allowing to confront the trans-Planckian 
question of inflationary cosmology and black hole physics. 

Third, we have applied our dissipative models to inflationary cosmology.
At early times, hence at very high energy, 
 the dissipative effects are so strong 
that all information about the initial state of the $\phi$ is erased,
see eq. (\ref{erase}).
Nevertheless, when the UV scale $\llv$ is much larger than the Hubble 
parameter, we have demonstrated that 
the standard expression of the power spectrum is found when 
the environment is in its ground state. The reason for this
 is the following: even though the field oscillator $\phi_p$ is
purely driven by ${\bf \Psi}_p$, i.e. it is 
given by the second term eq. (\ref{trues}), 
as its proper momentum $p/a$ redshifts under $\llv$, 
the composite operator behaves as if it were a free operator,
see eqs. (\ref{i1}, \ref{mie}), thereby 
guaranteeing eq. (\ref{dfree}).   

Let us also note that our models can be used for phenomenological purposes
in the sense of \cite{Jacobson:2005bg},
and that we are planning to apply them to study Hawking radiation
in the presence of dissipation.
We are presently completing the calculation 
of the power spectrum beyond the adiabatic approximation so as to 
determine the signatures of dissipation~\cite{AdamekCNP}
and to compare them to those of dispersion~\cite{Macher:2008yq}. 
Finally it would be interesting to compute the VEV
of the 
stress-tensor of the $\bf \Psi$ fields in non-trivial backgrounds. 
This would allow to take into account the back-reaction on the cosmological metric 
engendered by the fields $\bf \Psi$
and $l^\mu$~\cite{Jacobson:2000gw,Lemoine:2001ar,Shankaranarayanan:2005cs}.

%
%

\vskip .5 truecm
      
{\bf Acknowledgments.}    

\noindent
I am grateful to Dani Arteaga and Enric Verdaguer for common work
allowing me to deepen my understanding of dissipative effects. 
I am also grateful to  Julian Adamek, David Campo,
Ted Jacobson,  Jean Macher, Jens Niemeyer, and Valeri Rubakov
for interesting discussions.  
I wish to thank the organizers of the workshops on 
"Micro and Macro structure of spacetime" held in
Peyresq in June 2005, 2006, and 2007 where this work has been presented and discussed.
This work has been supported by the Agence Nationale de la Recherche
(projet 05-1-41810).

  
 \bigskip

\section{Appendix A : \\ Stationary states 
and vacuum 2pt functions}

   In this Appendix, we recall the (well-known) 
  relationships between $G_c$, $G_a$ which always hold
  in stationary states. 
In these states,
$G_c, \, G_a$ 
and the kernels $D, \, N$
are related to each other in a universal 
way,
generally referred as a Fluctuation-Dissipation relation.
We explain its origin and its physical implications
in the present context. We start 
with the most basic object: the retarded Green function $G_r$.

  
  \subsection{The retarded Green function}
  
  The Fourier transform of eq. (\ref{eqtosimpl})
  gives
  \be
   ( -\om^2 + \om_p^2 ) \phi_{\bf p}(\om) = 
  \Sigma_i\,   g_i(p) 
  \Psi_i^o({\bf p}, \om)
  + \Sigma_i\,   g_i^2(p) 
  R_i^o(\om; p)
    \phi_{\bf p}(\om) \, ,
    \ee
   where
   \be
   R_i^o(\om; p)= \left(-(\om + i \e)^2 +\Om_i^2(p))   \right)^{-1} \, ,
    \label{Grfpsi}
    \ee
    is the Fourier transform (defined as in eq. (\ref{FTdef})) 
    of the retarded Green function of $\Psi_i$.
   As usual, its retarded character is enforced by the
   imaginary prescription of the two poles to lay
   in the lower half plane ($\e > 0$).  
   The solution of the above equation is
   \be
    \phi_{\bf p}(\om) =   \phi_{\bf p}^d(\om) + G_r(\om, p)
  \Sigma_i\,   g_i(p) 
  \Psi_i^o({\bf p}, \om) 
  \label{truef}
     \, , 
    \ee
  where the  Fourier transform of the retarded Green function 
  of $\phi$, the solution of eq. (\ref{Gr}), always takes the form
  \be
   G_r(\om, p)  = \left( -(\om + i \e)^2 +p^2 + \Sigma_r(\om,p)
   \right)^{-1}
   \, .
   \label{Grf}
    \ee
  All effects of the coupling to the $\Psi_i$'s are thus encoded
  in the (retarded) self-energy $\Sigma_r(\om,p)$. For Gaussian theories, 
  it is {\it algebraically} given by
  \be
  \Sigma_r(\om,p)= - \Sigma_i \,  g_i^2(p) R_i^o(\om, p)\, .
  \label{Sr}
  \ee
  The dissipative effects are governed by the imaginary 
  part of $ \Sigma_r(\om,p)$.
  In the present case, one has
  \be 
2{\rm Im} \Sigma_r(\om,p) = - \Sigma_i \, g_i^2(p) G^o_{c, i}(\om, p) = - D(\om, p)\, .
\label{DSr}
\ee
To get the first equality we have used the fact that 
in stationary states the retarded Green function and the commutator 
are related by $2$Im$G_r(\om) = G_c(\om)$ for all degrees of freedom, 
free or interacting. In the second equality, we have introduced $D(\om)$,
the Fourier transform of the kernel of eq. (\ref{d}).
 
Several observations should be made here. 
First, from eq. (\ref{Grfpsi}), we obtain that $D(\om)$ is proportional to
$
\Sigma_i\,  g_i^2 \delta(\om - \Omega_i)$. Therefore there is no dissipation
for lower frequencies than the lowest value of $\Omega_i$. 
This simply follows from energy conservation. 
Second, to
obtain "true" dissipation, $D(\om, p)$ should be a continuous
function and not a sum of delta. 
  This can only happen
  when the $\Psi_i$
  form a dense ensemble.
In Section \ref{TDS}, 
we shall thus replace the discrete sum on $i$ by an integral
  over a 
  continuous variable, $k$. 
  We shall not consider the discrete cases
even though these could display interesting properties.  
 Third, from a phenomenological point of view, only $D(\om, p)$ matters. 
 Hence we cannot disentangle 
 the spectrum of the environment, which is given
 by $R_i^o(\om, p)$, from 
 the coupling strength $g_i^2(p)$.
 This is a good thing, because 
 when working in time-dependent settings, we shall exploit this equivalence
 to chose the simplest model of $\Psi_i$'s which gives the 
 kernel $D(\om, p)$ we want.


It is also worth noticing that the dispersive (real) effects
are not directly related to $D$ (or $N$). These 
are governed by the even part of $ \Sigma_r(\om,p)$ which is given by
\be 
{\rm Re}
\Sigma_r(\om,p) = \int\!  \frac{ d\om'}{2 \pi } 
\frac{ D(\om', p)}{\om - \om'}\,,
\label{Kra}
\ee
 where the integral should be understood as a principal value. 
This 
integral relationship
explains why one often founds that 
dispersive effects appear before dissipative effects (for increasing $\om$).
We also learn that the dispersive models studied in the literature 
violate the above relation since they assume 
Re$\Sigma_r \neq 0$ and 
 ${\rm Im} \Sigma_r \equiv 0 $. Therefore these models
 are incoherent and cannot 
  result from dynamical processes. 

  
  \subsection{Fluctuation-Dissipation relations and vacuum self-energy}
  
In this subsection, we derive the
 relationships between $G_c$, $G_a$ and $\Sigma_F$ which
 exist in the true (interacting) ground state.

In interacting theories, the only stationary states are thermal
states, see e.g. \cite{Anglin:1992uq}. 
In these states, 
the Fourier transform
of $D$ and $N$ are related by
\ba
N(\om) &=& D(\om)\, 
\coth(\beta \om/ 2) \, , \nonumber\\
&=& D(\om)\, {\sign(\om)} \, [2n(\vert \om \vert) + 1 ] \,  .
\label{fdur}
\ea
In the second line, $n(\om)$ is the Planck distribution. It gives
the mean occupation number of $\Psi^o_i$ quanta
as a function of the frequency (measured in the rest frame of the bath).
The above relation directly follows from the fact that
the individual commutators and anti-commutators
of the free fields $\Psi^o_i$ 
obey this relation, as any free oscillator 
would do. 
It implies that
the Fourier transform of 
$G_c$ and $G_a$ are also related by
\be
G_a(\om) = G_c(\om)\, {\sign(\om)} \,[2n(\om) + 1 ] \,  .
\label{fd}
\ee
It should be stressed that this equation is exact, 
i.e. non-perturbative, and valid for all theories, 
Gaussian or not. (It indeed directly follows from the cyclic properties of 
the trace defining $G_{\beta}(t,t')=$Tr$ [e^{-\beta H_T} \phi(t) \phi(t')]$). 
 
For Gaussian models, there exists an alternative proof 
of eq. (\ref{fd}). 
It suffices to note that 
in steady states the decaying terms of eqs. (\ref{Gcgs})
and (\ref{Gags}) 
play no role, and that 
the Fourier transform of the driven terms are  
respectively given by
\ba
G_c(\om) &=& \vert G_r(\om)\vert^2 \,  D(\om) \, , 
 \label{Gcf}
 \\
G_a(\om) &=& \vert G_r(\om)\vert^2 \, N(\om) \, ,
\label{Gaf}
\ea
since the Fourier transform of the
retarded Green function obeys $G_r(\om)=G_r^*(-\om)$, see eq. (\ref{truef}).
Irrespectively of the 
complexity of $G_r$, i.e. irrespectively of the functions $g_i(p)$, 
$\Omega_i(p)$ and the set of the $\Psi_i$
fields, 
$G_c$ and $G_a$ are thus related to each other
by the FD relation (\ref{fd}).


These universal relations will be relevant 
for inflationary models wherein only the ground state contributes.  
In particular, they
imply that {\it 
in the true vacuum}, i.e.
when $n(\om)= 0$,
$G_c$ and $G_a$ are exactly related by
$G_a(\om) = G_c(\om)\,\sign(\om) $. Hence the
Wightman function 
\be
 G_W = {1 \over 2} ( G_c + G_a) = G_c \, \theta(\om)\, ,
 \label{Gwf}
\ee
is determined by the commutator and contains only positive frequency, 
as in the free vacuum. 
Equations (\ref{Gcf},\ \ref{Gaf}) also
allow to compute the 
vacuum self-energy 
of the Feynman Green function. 
For
Gaussian models it is  given by
\be
2 {\rm Im}  \Sigma_F(\om, p) = - D(\om, p) \, {\sign(\om)} 
\, . 
\label{SfD}
\ee
With the last equality we recover the fact that in the vacuum, it is sufficient
to consider Feynman Green functions. In non-vacuum states, and in 
non-stationary situations, this is no longer true, thereby justifying
the use of the $in-in$ machinery~\cite{Arteaga:2007us} 
(the Schwinger-Keldish formalism).

Before specializing to a specific class of models 
giving rise to dissipation at high frequency, 
we make a pause by asking the following important question:
What should be known 
about the $\Psi_i$ fields to get eqs. (\ref{Gcf}, \ref{Gaf}, \ref{Gwf}) ?
We have proven that it is sufficient for the 
$\Psi_i$'s to be canonical fields, but is it necessary ?

The answer is two fold. 
On one hand, the $\Psi_i$ cannot be stochastically fluctuating quantities
(i.e. commuting variables) because this would lead
to a violation of eq. (\ref{fdur}) that would imply the
violation of 
eq. (\ref{fd}) and the ETC eq. (\ref{i1}).\footnote{
This constitutes the simplest proof that it is 
inconsistent to couple quantum variables to stochastic (or classical)
ones. If one does so, the ETC of the dressed quantum
variables will always be dissipated after a time of the order
of $\gamma^{-1}$. One can therefore view the experimental
evidences for the ETC of some degrees of freedom 
as a very strong indication that {\it all} 
dynamical variables in our world are quantum mechanical in nature.}
They cannot be either a combination of quantum and
stochastic quantities because this would still lead
to a violation of the ETC.
Hence they must be built only from quantum (canonical)
degrees of freedom.

On the other hand, the $\Psi_i$'s can be composite 
operators,
 i.e. polynomials of some 
(unknown) canonical fields.
Indeed, 
their commutators would still be all related
to their anti-commutators by the 
FD relation eq. (\ref{fdur}), and this even though  
they depend non-linearly on $n(\om)$ in non-vacuum states.  
The difference with Gaussian models is that 
these non-linear operators posses
non-vanishing higher order correlation
functions. Hence, 
the self-energies $\Sigma_r, \, \Sigma_F$
will be series in powers of 
$g_i^2$, and not just a single quadratic term as in
eq. (\ref{Sr}, 
\ref{SfD}).
Nevertheless 
these higher loops corrections 
preserve
the validity of eq. (\ref{Gwf}) in the ground state, 
 as well as that of eqs. (\ref{Gcf}, \ref{Gaf}) in any thermal state, when
properly understood, i.e. with $D$ now defined by -2Im$\Sigma_r$ (as the 
effective dissipation kernel), and $N$
related to it by the FD relation. 


In brief, we have reached/recalled the following 
results.
Firstly, 
the q-number combination ${\bf \Psi}=\Sigma_i\,  g_i\Psi_i$, 
the fluctuating source term of $\phi$, must obey 
the 
FD relation (\ref{fdur}).
This can either be postulated, 
or better, 
be viewed as resulting from the fact that
${\bf \Psi}$ is entirely made out of 
quantum 
degrees of freedom.
Secondly, to lowest order term in $g$, the self-energy
can be obtained by treating ${\bf \Psi}$
as a 
Gaussian variable, what ever its composition may be. 
Thirdly, when dealing with non-Gaussian theories, once 
having computed $\Sigma_r(\om)$, the resulting equations
for the 2pt functions have the same structure and the
same meaning as in Gaussian theories, with $D$ replaced
by -2Im$\Sigma_r$. Therefore, {\it the phenomenology of 
two-point functions is entirely covered by 
Gaussian settings}.

\subsection{The double limit: $g^2 T \to \infty$ followed by $g^2 \to 0$.}

To perform a phenomenological analysis of dissipation, 
 we need to 
 understand
 how the theory behaves in transitory regime from dissipative 
 to free propagation.
Similarly, 
to study primordial spectra in inflation or Hawking radiation,
we also need to understand how free motion emerges as the 
proper frequency get red-shifted. 
It is therefore relevant 
to study the 
behavior of the two-point function in the 
following double limit. 

One first takes $g^2 T \to \infty$, 
where $T = t-t_{in}$,
$t_{in}$ being 
the moment when the interactions are turn on, and $t$ the moment when 
one studies the field properties. 
This limit implies that the decaying term in 
eq. (\ref{trues}) plays no role. Therefore, near time $t$, 
 the Heisenberg field $\phi(t)$ is a composite operator 
 which only acts in the Hilbert space of $\bf \Psi$.
 
 Secondly, one considers the "free" limit $g^2 \to 0$ of that composite operator. 
 One could naively conclude that $G_c$ and $G_a$ of eqs. (\ref{Gcf},\, \ref{Gaf})  
 would vanish since
 both $D$ and $N$ are proportional to $g^2$. However, this is not the case, 
 because the common prefactor, 
 $\vert G_r \vert^2$,
 is singular in this second limit. In fact, one verifies that it scales in
 $1/g^2$ in such a way that, in the (interacting) vacuum, one always recovers
 \be
 G_W(\om)_{g^2 \to 0} = 
 {1 \over 2 \om_p}\, 2\pi \delta(\om - \om_p 
  )\, .
 \label{mie}
 \ee
 This 
 is the standard vacuum fluctuations of a free massless mode of 
 momentum $p$. 
 
 Two important lessons have been reached. 
 First we learned
 is that even though $\phi$ acts 
 {\it only} on the $\bf \Psi$-Hilbert space, 
 when $g^2 \to 0$, it behaves as if it were a free mode 
 possessing its own Hilbert space, with no reference to $\bf \Psi$-dynamics.
 Secondly, 
 the quantum state in this would be Hilbert space
  is 
  still exactly that of $\bf \Psi$.
Therefore, in stationary situations, the only "souvenir" kept by the composite 
operator
is the equilibrium distribution $n(\om)$ inherited from its parents. 

 Let us now emphasize that the above limit is relevant for non-Gaussian models
 as well. 
 Indeed, in the limit $g^2 \to 0$, there will always be a value of $g^2$
 sufficiently small that the model can be well approximated by a Gaussian
 model. Therefore the 
 behavior of the 2-point functions in the transitory regime 
 from dissipation to free
 propagation can be analyzed 
 by studying Gaussian models (at least in the quasi-static limit). 

\section{Appendix B: \\
Dissipative effects above $\llv$. The Phenomenology}

 We now have all the tools to construct 
 models giving rise to dissipation in the vacuum
 above a critical energy scale $\llv$. 
 In this Appendix we work from a purely phenomenological point of view,
 and provide the class of dissipative models wherein
the imaginary part of the self-energy is governed by a single term, 
in analogy with the dispersion relations of eq. (\ref{disprel}). 
%

From a 
  phenomenological point of view,
  if one considers only stationary situations (i.e. static metrics 
  and stationary states), 
   one   can simply choose the function $D(\om,p)$ entering
 eq. (\ref{DSr})  and  eq. (\ref{Gcf}) {\it as one wishes}. There is indeed
  no restriction on $D(\om,p)$ besides its constitutive properties, namely 
  being odd in $\om$ and giving rise
  to poles in $G_r$ all localized in the lower half $\om$ plane. 
  In this we have reached our first aim, namely
  identify how to generalize the free settings so as to 
  incorporate some arbitrary dissipative effects.

  We can thus consider the dispersive models which correspond to 
those defined by  
eq. (\ref{disprel}). They 
 are characterized by a 
  single term giving rise to dissipation above $\llv$. 
  In the vacuum, they are 
fully specified 
 by the 
imaginary part of the (retarded) self-energy 
  \be
 -{\rm Im}\Sigma^{(n)}_r(\om, p) 
    = {\om \over \llv}
    \, p^2 
   \left({p \over \llv}\right)^{2n}
   = 2 {\om} 
   \, \gamma_{n}\, . 
   \label{iDRn2}
   \ee
   In these models, 
   the decay rate (inverse life time) on the mass shell
   is 
   \be
  { 
  \gamma_{n}} =
  {p \over 2} \left({p \over \llv}\right)^{2n+1} \, .
  \label{decayn} 
  \ee
   To verify it, assuming 
  that Re$\Sigma_r = 0$,
  the two poles of $G_r(\om)$ in eq. (\ref{Grf}) 
  are located in
   \be
   \om_\pm(p) = \pm \sqrt{\om_p^2 - \gamma^2} - i \gamma
   \label{omeff}
   \, .
   \ee
   From this, by inverse Fourier transform $G_r(\om)$,  
    one obtains that the
   decay rate 
   is indeed $\gamma$ in the under-damped regime, for $\om_p^2 > \gamma^2$.
  In the overdamped regime, for $\gamma^2 > \om_p^2 $, the decay rates of the 
  two independent solutions of  $G_r^{-1} \phi_d = 0$
  are $\Gamma_\pm = \gamma \pm \sqrt{\gamma^2 - \om_p^2}$.
  
   One thus have the following behavior as $p$ grows.  
   For $p \ll \llv$, $\om_\pm  \simeq p$, and one has
    a free propagation which is slightly damped 
    with a life time in the units of the frequency given by  
   $(\llv/\om)^{n+1} \gg 1$. 
   In the opposite regime of high momenta $p \gg \llv$, deep in the overdamped
   regime, the two roots $\om_\pm$ are real
   and the notion of propagation (in space-time) is  absent. 
   In anticipation to what will occur in inflation (or in black hole physics),
   we invite the reader 
   to study the migration of the poles of $G_r$ 
   when extrapolating backwards in time a mode, 
   i.e. as $p$ increases. (Remember that the physical momentum of a mode 
   in cosmology is $p_{phys}(t) \propto p/a(t)$ where $p$
   is the norm of the conserved comoving wave vector
   (near a black hole one finds $p(r)\propto \om /x$ where $x = r-r_S$
   is the proper distance from the horizon measured in a freely falling frame,
   and $\om$ the conserved Killing frequency.))

   One could of course generalize the above class
   by considering in eq. (\ref{iDRn2})
   polynomials in $p$ dimensionalized by different UV
    scales. However, unless fine tuning, the phenomenology 
    of the transition from the IR dissipation-free sector to the
    dissipative sector will be dominated a single term. One should also 
    consider the possibility that Im$\Sigma$ strictly vanishes 
    below a certain frequency $\Omega_1$,
    as this would be the case when ever the spectrum of the $\Psi$ fields
    possesses this gap, see the remarks after eq. (\ref{DSr}).
    
 Having the phenomenology of dissipative and unitary models under control
 (with dispersive and dissipative effects related by Kramers relation, 
  eq. (\ref{Kra}))
 one could confront particle and astro-particle 
    physics data and put lower bounds on $\llv$ for each $n$, in analogy with what 
    was done for (pure) dispersion in \cite{Jacobson:2005bg}. 


\end{document}